\documentclass[fleqn,usenatbib]{mnras}

\usepackage{newtxtext,newtxmath}

\usepackage[T1]{fontenc}
\usepackage{lscape}
\DeclareRobustCommand{\VAN}[3]{#2}
\let\VANthebibliography\thebibliography
\def\thebibliography{\DeclareRobustCommand{\VAN}[3]{##3}\VANthebibliography}

\usepackage{graphicx}	
\usepackage{amsmath}	
\usepackage{caption}
\usepackage{subcaption}
\usepackage{tikz}
\usepackage{float}
\usepackage{multirow}
\usepackage{lipsum}  
\usepackage{color,soul}
\usepackage{ulem}
\usepackage[]{mdframed}

\definecolor{latexgreen}{rgb}{0.74902, 0.86667, 0.75294}
\definecolor{lemonchiffon}{rgb}{1.0, 0.98, 0.8}
\definecolor{flax}{rgb}{0.93, 0.86, 0.51}
\definecolor{blizzardblue}{rgb}{0.67, 0.9, 0.93}
\definecolor{cadetgrey}{rgb}{0.57, 0.64, 0.69}
\definecolor{cambridgeblue}{rgb}{0.64, 0.76, 0.68}
\definecolor{cardinal}{rgb}{0.77, 0.12, 0.23}
\definecolor{paleaqua}{rgb}{0.74, 0.83, 0.9}

\setul{0.5ex}{0.4ex}
\setulcolor{latexgreen}

\setstcolor{red}

\def\code#1{\texttt{#1}}

\DeclareRobustCommand{\rchi}{{\mathpalette\irchi\relax}}
\newcommand{\irchi}[2]{\raisebox{\depth}{$#1\chi$}}

\title[Chemical enrichment in M89]{The infalling elliptical galaxy M89: The chemical composition of the AGN disturbed hot atmosphere}

\author[S. Kara et al.]{
Sinancan Kara,$^{1}$\thanks{E-mail: kara.sinancan@gmail.com}
Tomáš Plšek,$^{2}$
Klaudia Protušová,$^{2}$
Jean-Paul Breuer,$^{2}$
Norbert Werner,$^{2}$ \newauthor
Fran\c{c}ois Mernier,$^{3,4}$ 
E. Nihal Ercan$^{1}$
\\
$^{1}$ Department of Physics, Bo\u{g}aziçi University, Bebek, 34342 Istanbul, Turkey\\
$^{2}$ Department of Theoretical Physics and Astrophysics, Faculty of Science, Kotlá\v{r}ská 2, Masaryk University, Brno, 611 37, Czech Republic\\
$^{3}$ NASA Goddard Space Flight Center, 8800 Greenbelt Rd, Greenbelt, MD 20771, USA \\
$^{3}$ Department of Astronomy, University of Maryland, College Park, MD 20742, USA}

\date{Accepted 2024 January 5. Received 2023 December 22; in original form 2023 September 12}

\pubyear{2022}

\begin{document}
\label{firstpage}
\pagerange{\pageref{firstpage}--\pageref{lastpage}}
\maketitle

\begin{abstract}
The chemical enrichment of X-ray-emitting hot atmospheres has hitherto been primarily studied in galaxy clusters. These studies revealed relative abundances of heavy elements that are remarkably similar to Solar. Here, we present measurements of the metal content of M89 (NGC\,4552), an elliptical galaxy infalling into the Virgo cluster with a $\sim$10 kpc ram-pressure stripped X-ray tail. We take advantage of deep {\it Chandra} and {\it XMM-Newton} observations, and with particular attention to carefully modelling the spectra, we measure the O/Fe, Ne/Fe, Mg/Fe, Si/Fe and S/Fe ratios. Contrary to previous measurements in galaxy clusters, our results for the hot atmosphere of M89 suggest super-Solar abundance ratios with respect to iron (i.e. $\alpha$/Fe > 1), similar to its stellar components.
Our analysis of the active galactic nucleus (AGN) activity in this system indicates that the AGN-induced outflow could have facilitated the stripping of the original galactic atmosphere, which has been replaced with fresh stellar mass loss material with super-Solar $\alpha$/Fe abundance ratios.
Additionally, we report a new fitting bias in the RGS data of low-temperature plasma. The measured O/Fe ratios are >1$\sigma$ lower in multi-temperature models than a single temperature fit, leading to discrepancies in the calculations of supernova fractions derived from the metal abundances.
\end{abstract}




\section{Introduction}\label{sec:introduction}

Massive galaxies, galaxy groups and galaxy clusters are pervaded by hot, X-ray-emitting diffuse gas, which contains 80\% of the total baryonic matter of the Universe (for a recent review, see e.g. \citealp{Werner_2020}). Since the $\sim$7 keV K-shell Fe emission lines were discovered in intracluster medium (ICM) X-ray spectra in the late 1970s \citep{Mitchell1976, Serlemitsos1977}, the hot plasma pervading the Universe is known to be enriched with heavy elements.


Through decades, it has been debated if the main enrichment mechanisms of the hot atmospheres in such systems are internal (e.g. stellar winds at galactic scale) or external (gas inflows from the intergalactic medium). Measurements of heavy elements play a crucial role in answering this question. The observations show that the spatial metal distribution does not follow the galaxy distribution, and the metal distribution in cluster outskirts is remarkably uniform \citep{Werner2013, Urban2017}, indicating significantly homogeneous gas formation. Furthermore, there is no apparent dependency of metal abundances on the mass of the galaxy cluster \citep{dePlaa2017, Mernier2018m, Truong2019}. In line with the observations, simulations indicate \citep{Biffi2017} that the enrichment of the ICM was completed at $z \sim 2-3$ ($10-12$ billion years ago), which is the epoch of peak star formation and active galactic nuclei (AGN) activity \citep{MadauDickinson2014, Hickox2018}. 

The chemical enrichment of the Universe on the largest scales occurred at early times when energetic AGN feedback outflows expelled the metals to the intergalactic medium. Later, the enriched and mixed atmosphere gas was externally accreted and heated by clusters, galaxy groups, and galaxies to form the present atmospheres (for a recent review, see e.g. \citealp{Mernier2022}).

Measurements of abundance ratios of light $\alpha$ elements (e.g. O, Ne, Mg, Si, and S), mostly produced by core-collapse supernovae (SNcc), and heavy Fe-peak elements (e.g. Ca, Cr, Mn, Fe and, Ni), mainly produced by Type Ia supernovae (SNIa), can be used to understand the chemical enrichment histories of different systems. \cite{Mernier2016a} analysed central chemical composition of 44 systems using \textit{XMM-Newton} EPIC and RGS observations. They found the chemical composition in their samples to be very similar to our Solar system (i.e. $\alpha$/Fe ratio is $\sim$1 Solar). Similarly, by utilising the high-resolution \textit{Hitomi} SXS and \textit{XMM-Newton} RGS observations, \cite{Simionescu2019b} found Solar abundance ratios in the Perseus cluster. These and other studies \citep[e.g. ][]{Hitomi2017, Grandi2009} indicate that the relative abundances of supernova products have Solar ratios from cores to the outskirts. These remarkable discoveries strongly suggest that the hot plasma pervading our Universe has become chemically well-mixed, achieving Solar composition through continuous inflows and outflows.

In cluster galaxies, the chemical composition of hot atmospheres and the stellar population are considered to be decoupled. The externally accreted hot gas has Solar abundance ratios (i.e. $\alpha/$Fe $\approx$ 1), while the stars have super-Solar abundance ratios (i.e. $\alpha/$Fe $>$ 1) \mbox{\citep{Conroy_2013}}. The origin of the decoupling is that the elliptical (i.e. early-type) galaxies are considered to have had rapid and intense star formation very early in cosmic history, with a star formation peak around $z \sim 3$ \mbox{\citep{Thomas2010}}. As a result, considering SNcc products formed earlier in the cosmic time, SNcc products have been locked in stars before the star-forming gas in such a galaxy has been polluted by SNIa products that are dominantly produced later in the cosmic time.

Although the hot galaxy atmospheres consist mainly of externally accreted and heated gas, internal sources also produce a large amount of hot gas. Internal sources of hot gas in elliptical galaxies are the thermalised stellar mass loss products \mbox{\citep{Mathews1990, MathewsBrighetni2003}}. In a case without any external inflow of Solar gas, stellar wind-originated gas with super-Solar abundance ratios might dominate the galaxy's atmosphere over time. However, \mbox{\cite{Mernier2022b}} recently showed that NGC\,1404, an infalling elliptical galaxy experiencing such an accretion cut-off, has Solar abundance ratios, which means internally produced hot gas has not dominated the overall composition in that system. Still, in the current picture of hot gas replenishment in elliptical galaxies, enrichment processes have not been tested in a system that might also undergo a significant loss of its original atmosphere due to internal mechanisms like AGN activity.

\begin{figure*}
    \centering
    \begin{tikzpicture}[remember picture]
        \draw (-6.75, 0) node {\includegraphics[height=185pt]{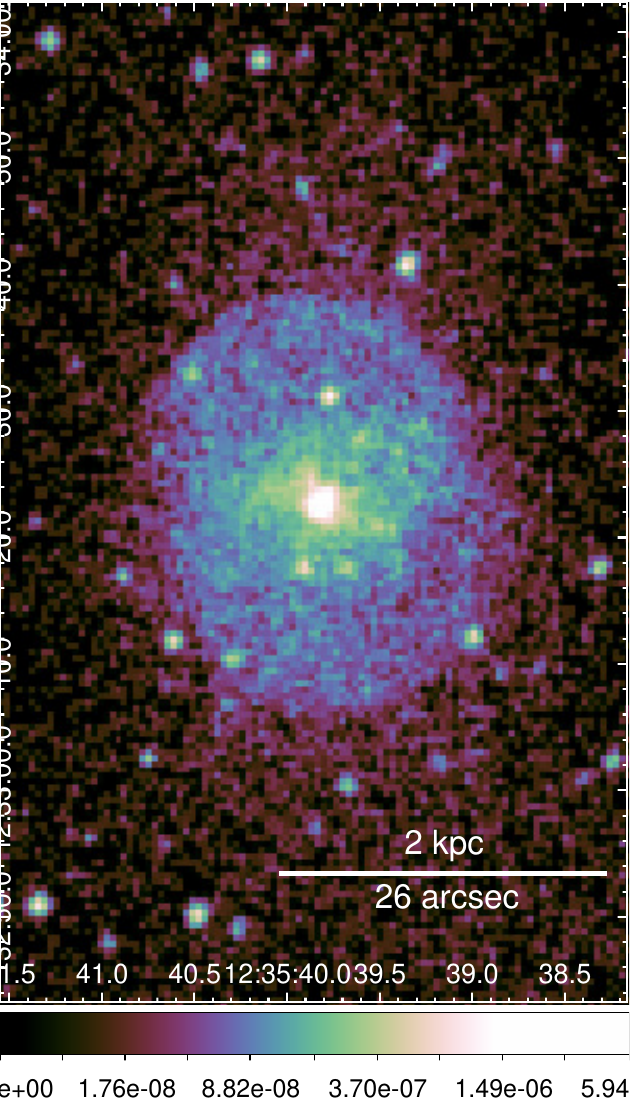}};
        \draw (-0.2, 0) node {\includegraphics[height=185pt]{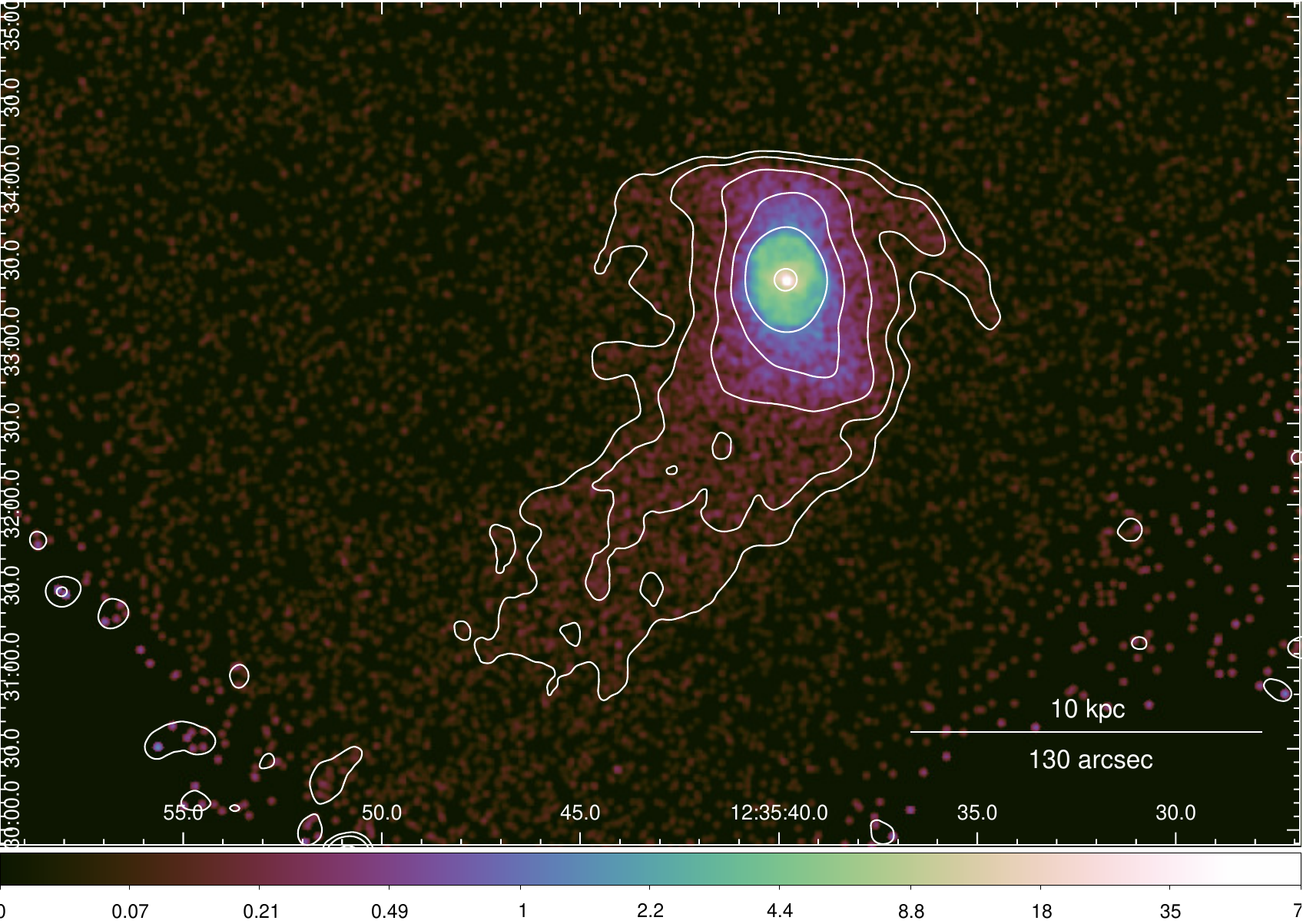}};
        \draw (6.75, 0) node {\includegraphics[height=185pt]{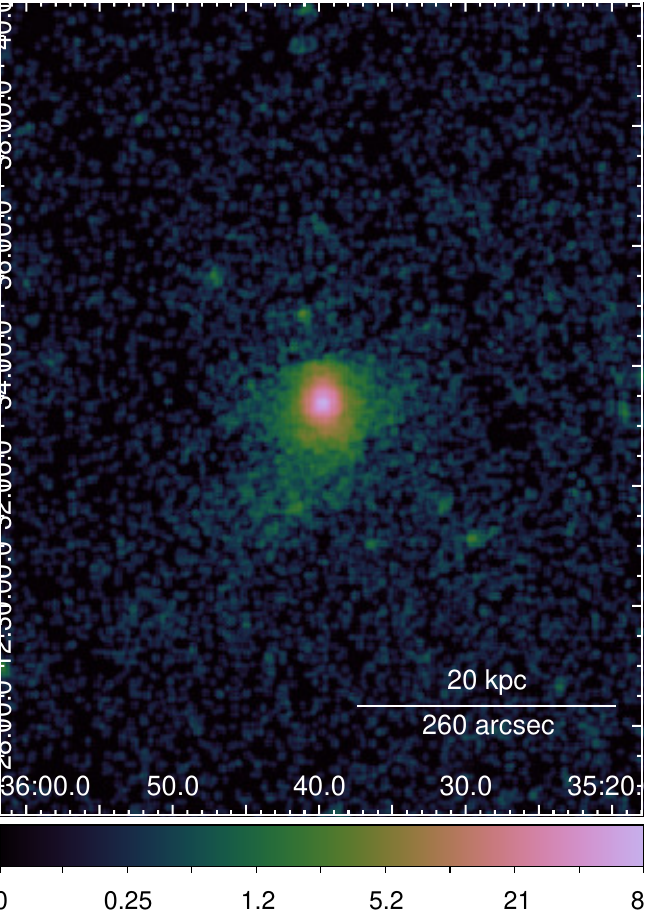}};
    \end{tikzpicture}
    \caption{(\textit{Left}) Exposure corrected, \textit{Chandra}/ACIS-S mosaic image of the core of M89 in $0.5-2.0$ keV band. Two ring-shaped AGN-induced shock rims are present. The point sources are left for visualisation purposes. (\textit{Middle}) Point source removed \textit{Chandra}/ACIS-S mosaic image of M89 and the Ram-pressure associated tail in the $0.7-1.1$ keV band, with the corresponding white contours of X-ray surface brightness. (\textit{Right}) Exposure corrected \textit{XMM-Newton}/EPIC image of M89 in $0.7-1.1$ keV band.}
    \label{fig:image_analysis}
\end{figure*}

M89 (a.k.a. NGC\,4552) is an excellent target to examine the hot gas replenishment processes due to stellar mass loss products. M89 is a massive elliptical galaxy experiencing AGN activity, oscillating inside the gravitational potential well of the Virgo cluster. Because of its motion, any gas inflow with Solar abundance ratios onto the galaxy is stopped \citep{Gunn_1972}. Moreover, due to the AGN activity, it is possible that the galaxy might have lost a portion of its original atmosphere with Solar abundance ratios. Therefore, due to its exceptional dynamics, a dedicated enrichment study on M89 has the potential to provide new constraints on the chemical enrichment of the Universe. The galaxy is located 350 kpc (72 arcmin) east of the central dominant galaxy of the cluster (NGC\,4486, a.k.a. M87). \textit{Chandra} images in the 0.5-2.0 keV band (Figure \ref{fig:image_analysis}) show two ring-like structures approximately $\sim$1.3 kpc away from the centre of the galaxy, which is associated with the LINER type AGN in its centre \citep{Machacek_2006b}. Moreover, it shows a $\sim$10 kpc X-ray tail at its south, along the motion direction, associated with the ram-pressure stripping of the galaxy's hot atmosphere \citep{Machacek_2006a}. Notably, the $\alpha$/Fe ratios in the stellar content of M89 are higher than the average ratios in ellipticals with similar mass. \citet{Lonoce2021} measured the abundance values within 0.975$r_e$ with 10 symmetric galactocentric bands, where the 1D extraction corresponds to $\sim$38.4 arcsec. From this study, stars in M89 have O/Fe ratios of $3.02_{-0.17}^{+0.15}$ and Mg/Fe ratios of $1.92_{-0.15}^{+0.16}$. 

The origin of the hot atmosphere of M89 can be investigated via the abundance ratios in the system. Previous abundance ratio measurements of M89 were first conducted by \cite{Ji2009}. They analysed M89 within a sample of 10 elliptical galaxies and found Solar abundance ratios for Mg/Fe and Si/Fe and sub-Solar for O/Fe in a circular region of 1 arcmin radius. However, their measurement assumes only a single-temperature model, which is known to result in unreliable measurements in cool systems ($kT \lesssim 2-3$ keV) due to the so-called `Fe bias' \citep{Buote1998, Buote2000, Gastaldello2021}. The `Fe bias' is an underestimation of Fe abundance when a spectrum containing multiple temperature components is estimated with a single-temperature model. Given that multi-temperature components are inevitable in the spectrum of a galaxy's atmosphere due to issues such as projection effects and cooling in the galaxy's core, we can assume that such measurements need to be revisited. Furthermore, \cite{Ji2009} reported unusually high abundances in M89, findings that were not corroborated by the CHEERS (the CHEmical Evolution RGS Sample) catalogue \citep{dePlaa__2017, Mernier2016a}. In the CHEERS study, they found super-Solar O/Fe and Si/Fe ratios and Solar Ne/Fe using single-temperature models. Although necessary to understand chemical enrichment processes as a whole, studies based on a sample of several systems at once are, by nature, much less suited to investigate individual outliers.


In this paper, we re-visit this elliptical galaxy with a detailed and comprehensive analysis. In particular, taking advantage of \textit{XMM-Newton} and \textit{Chandra} observations, we investigate the chemical enrichment of M89 for the first time by presenting the O/Fe, Ne/Fe, Mg/Fe, Si/Fe and S/Fe abundance ratios in the core and the tail of M89. Moreover, to comprehensively explore how the dynamical processes within M89 may influence its chemical enrichment history, we analyse the shock properties exhibited by the "hourglass" structure within M89, which is caused by the nuclear activity of the central AGN.

The structure of this paper is as follows. Section~\ref{sec:sec2_datareduction} describes the data reduction and region selection. The spectral modelling is described in Section~\ref{sec:sec_3_spectral_analysis}. Section~\ref{sec:Results} presents the abundance ratio results, supernova contribution fractions derived from the ratios, the systematic uncertainties in the measurement, and the properties of AGN-induced nuclear outburst. We discuss the possible chemical enrichment history of M89 in Section~\ref{sec:sec5_discussion}. Finally, Section~\ref{sec:sec6_conclusion} summarises the results and conclusions. Throughout the paper we assume the standard $\Lambda$CDM cosmology with $H_0 = 70\;$km$\,$s$^{-1}\,$Mpc$^{-1}$, $\Lambda_0 = 0.73$ and $q_0 = 0$. All the abundances are expressed using the proto-Solar values of \citet{Lodders2009}, and for simplicity, the values are referred to as `Solar' throughout the paper. Unless stated otherwise, all uncertainties are expressed in the $1\sigma$ credible interval\footnote{This corresponds to distances of $15.9\,\%$ and $84.1\,\%$ quantiles from the median value.}.




\setlength{\tabcolsep}{5.0pt}
\renewcommand{\arraystretch}{1.25}
\begin{table*}
\caption{\textit{Chandra} and \textit{XMM-Newton} observations of M89. Both \textit{XMM-Newton} RGS and EPIC data are taken from the same single \textit{XMM-Newton} observation.}
\begin{tabular}{cccccc}
\hline
\multicolumn{1}{|c|}{Observatory} & \multicolumn{1}{c|}{\phantom{aaaaa}ObsID\phantom{aaaaa}} & \multicolumn{1}{c|}{Observation Date} & \multicolumn{1}{c|}{Instrument} & \multicolumn{1}{c|}{Total Exposure (ks)} & \multicolumn{1}{c|}{Clean Exposure (ks)} \\ \hline
\multirow{4}{3.5em}{\textit{Chandra}} & 2072  & 2001-04-22 & \multirow{4}{3.4em}{ACIS-S} & 54.4 & 54.2\\
                                      & 13985 & 2012-04-22 &                             & 49.4 & 49.4\\
                                      & 14358 & 2012-08-10 &                             & 49.4 & 49.4\\
                                      & 14359 & 2012-04-23 &                             & 48.1 & 48.1\\
\hline
\multirow{4}{5.7em}{\textit{XMM-Newton}} & \multirow{4}{4.5em}{0141570101} & \multirow{4}{4.8em}{2003-07-10} & EPIC MOS 1 & 42.7 & 24.0 \\
 & & & EPIC MOS 2 & 42.7 & 24.1 \\
 & & & EPIC pn & 39.5 & 17.1 \\
 & & & RGS & 43.7 & 24.3 \\
\hline
\end{tabular}
\label{table1}
\end{table*}

\section{Data Reduction and Region Selection}
\label{sec:sec2_datareduction}

\subsection{Chandra X-ray observatory}

For our study, we used all available archival \textit{Chandra} observations of M89 with a total cleaned exposure of 201\:ks (see Table \ref{table1}). All observations were performed using the Advanced CCD Imaging Spectrometer array with ACIS-S (chip S3) in the aim point. Since the desired area of M89 does not overlap with other chips, we limited the analysis to the S3 chip. All observations were processed using standard \code{CIAO} 4.15.1 \citep{Fruscione2006} procedures and current calibration files (CALDB 4.10.4).


\textit{Chandra} observations were reprocessed and filtered for VFAINT events using the \code{chandra\_repro} script and deflared using the \texttt{lc\_clean} algorithm within the \texttt{deflare} routine. Individual OBSIDs were further reprojected to the same tangent point. These reprojected event files were then merged. The images were extracted in the $0.7-1.1$ keV and $0.5-2.0$ keV bands and exposure-corrected using weighted exposure maps (the weighting was computed using \texttt{make\_instmap\_weights}\footnote{\href{https://cxc.cfa.harvard.edu/ciao/ahelp/make_instmap_weights.html}{https://cxc.cfa.harvard.edu/ciao/ahelp/make\_instmap\_weights.html}} procedure assuming an absorbed \texttt{vgadem} model with $kT \approx 0.64$ keV; see Table \ref{table:parameters}).

{\it Chandra} X-ray images of M89 in the $0.7-1.1$ keV and $0.5-2.0$ keV bands are shown in Figure \ref{fig:image_analysis}. The ram-pressure stripped tail of $\sim$10 kpc is clearly visible. 
In the central region, we can see the diffuse emission from the galaxy along with two ring-like structures associated with AGN-driven weak shocks. 

\textit{Chandra} spectral files were extracted using the \code{specextract} script. When extracting spectra, we used stowed spectral files obtained using the \texttt{blanksky} script for background subtraction for all extraction regions. The reason for choosing stowed spectral files rather than the blank-sky data set is because stowed files only account for the instrumental background and thus allow for more accurate modelling of the sky background components. This is particularly crucial for low-temperature background components that are comparable to the temperature structure of M89. Robustly constraining the background components is essential for accurate abundance measurements, as further explained in Section~\ref{section:background}. In Section~\ref{sec:Results}, we compare our results obtained from the stowed spectral files and blank-sky data set. The background files were reprojected onto the observations, filtered for VFAINT events, and the background spectra were scaled to match the particle background of the observations in the $9-12$ keV energy range.

\begin{figure}
    \centering
    \includegraphics[width=\columnwidth]{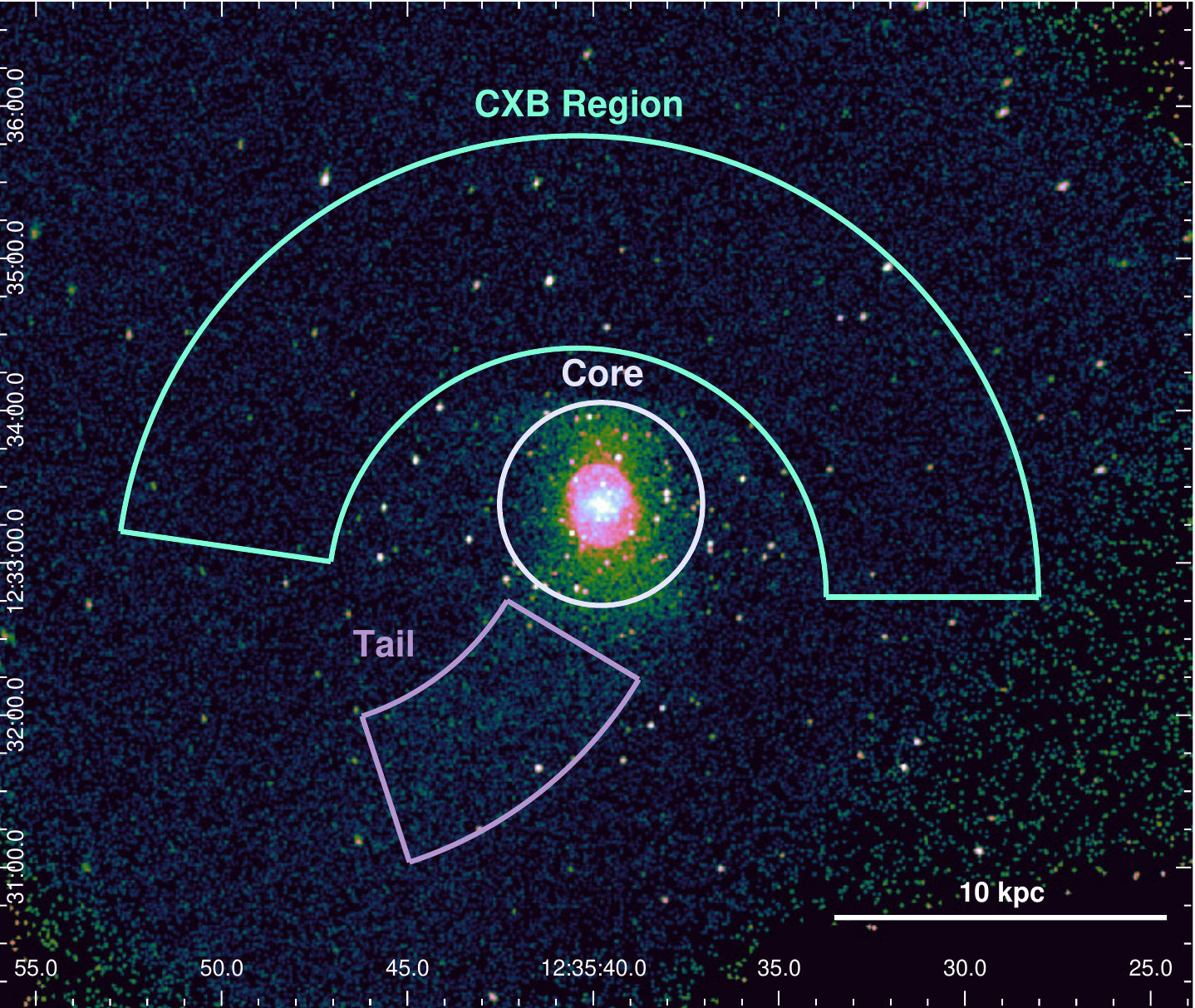}
    \includegraphics[width=\columnwidth]{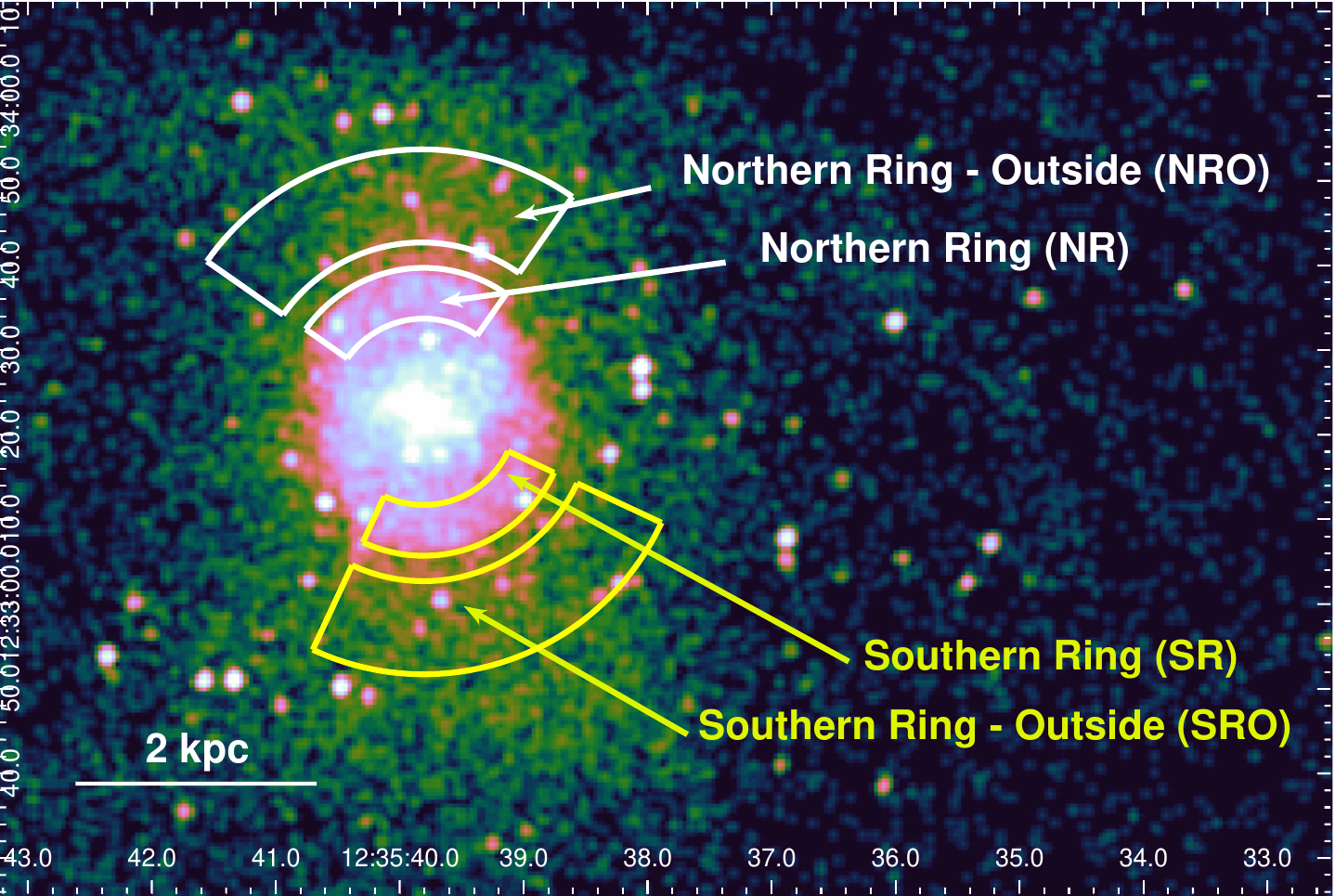}
    \caption{\textit{(Top)} Background subtracted \textit{Chandra}/ACIS-S image of M89 with highlighted regions used for our abundance study: Cosmic X-ray Background (CXB) region, Core region, and Tail region. \textit{(Bottom)} Zoom onto the centre of the galaxy with highlighted extraction regions used for estimating parameters of the shock.}
    \label{fig:regions}
\end{figure}

\subsection{XMM-Newton}

We used the single \textit{XMM-Newton} observation listed in Table \ref{table1}. Reduction of the EPIC and RGS data is done by using the \textit{XMM-Newton} Science Analysis System (\code{SAS} v20.0.0), along with the integrated Extended Source Analysis Software (\code{ESAS} v9.0\footnote{\href{https://www.cosmos.esa.int/web/xmm-newton/xmm-esas}{https://www.cosmos.esa.int/web/xmm-newton/xmm-esas}}) package.

\subsubsection{EPIC}

Following the standard procedure, the calibrated photon event files of MOS and pn data were created using \code{emchain} and \code{epchain}, respectively. Then, in order to exclude the soft proton (SP) contamination, we used \code{mos-filter} and \code{pn-filter}, which call the task \code{espfilt} to create good time intervals (GTI). We kept the single, double, triple, and quadruple events in the MOS data (\code{pattern$\leq$12}). Due to the charge transfer inefficiency problems for double events of pn, we kept the single events only (\code{pattern$=$0}). After the filtering, the exposure time has decreased from 44.3 ks to a net 24.3 ks due to the heavy contamination by flares. A combined, exposure-corrected image is merged with the \code{combimage} task and presented in Figure \ref{fig:image_analysis}. Point sources were detected and removed by the \code{CIAO} algorithm \code{wavdetect}. The Point-Spread Function (PSF) map of EPIC was created using the \code{dmimgcalc} tool for the purpose of the \code{wavdetect} algorithm. We generated corresponding PSF maps with a size of 9 arcsec, roughly equivalent to a 0.50 Encircled Energy Fraction (ECF) on-axis. After performing the point source detection, we see that the only resolved and detected point sources are located outside the central region of the galaxy. In Section \ref{sec:Results}, we discuss variations in results obtained from background spectra with different sizes of removed point sources.

We processed the \textit{XMM-Newton}-EPIC data by using \code{mos-spectra} and \code{pn-spectra} tasks, which call the \code{evselect}, \code{rmfgen} and \code{arfgen} tasks to extract the spectra and generate RMFs and ARFs, respectively.

\subsubsection{RGS}

We used the \code{SAS} task \code{rgsproc} to process the RGS data. The RGS1 and RGS2 data were filtered with the same GTI file for MOS1. To reduce the instrumental line broadening due to the slit-less nature of the gratings, we included 90\% of the PSF along the cross-dispersion direction (\code{xpsfincl=90}), which corresponds to a spatial width of $\sim$0.8 arcmin. We combined the RGS1 and RGS2 spectra with \code{rgsproc} task and fitted the first-order and second-order data simultaneously.

\subsection{Region Selection}
\label{region_selection}


The emission of the hot atmosphere of M89 is, apart from the cosmic X-ray background, highly contaminated also by the ICM emission of the Virgo cluster. In order to properly estimate the abundance of chemical elements in M89, it is critical to describe this background emission well. Furthermore, the central active galactic nucleus (AGN) can also contribute to the observed spectrum and can only be properly spatially excluded using \textit{Chandra} data. For these reasons, we adopted the following approach and extracted spectra from multiple regions.

First, to investigate the chemical composition of the disturbed hot gas of M89, we extracted a spectrum from a circular region at its centre with a 40 arcsec radius (Core region), presented in Figure \ref{fig:regions}. The size of this region is selected to correspond to the galaxy core and also, considering the relatively short clean exposure time of \textit{XMM-Newton}, to be large enough to have a sufficient number of counts so that the abundances can be constrained robustly. For \textit{Chandra} data, the circular region with a 1.5 arcsec radius corresponding to the central AGN was excluded. In order to investigate the elemental structure of the escaped gas, we also created a Tail region enclosing the galaxy's X-ray tail.

In order to model the Virgo ICM and the cosmic X-ray background (CXB) robustly, we created an arc-shaped region (CXB Region) in front of the galaxy along the direction of the motion of M89. The CXB region (see Figure \ref{fig:regions}) and the Core or Tail region are modelled simultaneously by linking the joint background parameters. The modelling procedure is further explained in Section~\ref{sec:sec_3_spectral_analysis}. 

To describe the time-variability and spectral properties of the central AGN, we extracted \textit{Chandra} spectra from a circular region with a 1.5 arcsec radius centred at the bright central point source (AGN Region).


Additionally, in order to examine the dynamics of the AGN-induced shock, we created region pairs encompassing each shock ring individually in annular sections with the inner and outer radii of 11 and 17 arcsec, respectively, while the outside regions of the rings are present in annular sections of inner and outer radii of 30 and 31 arcsec centred around the central AGN. For the northern ring, the annular sections Northern Ring (NR) and Northern Ring - Outside (NRO) regions extend from $55^{\circ}$ to $145^{\circ}$. The Southern Ring (SR) and Southern Ring - Outside (SRO) regions extend from $245^{\circ}$ to $335^{\circ}$ (see Figure \ref{fig:regions}). These regions are used to investigate the temperature discrepancy present across the rim of the shock, which enables us to determine the shock Mach number.




\begin{figure*}
    \centering
    \includegraphics[width=\textwidth]{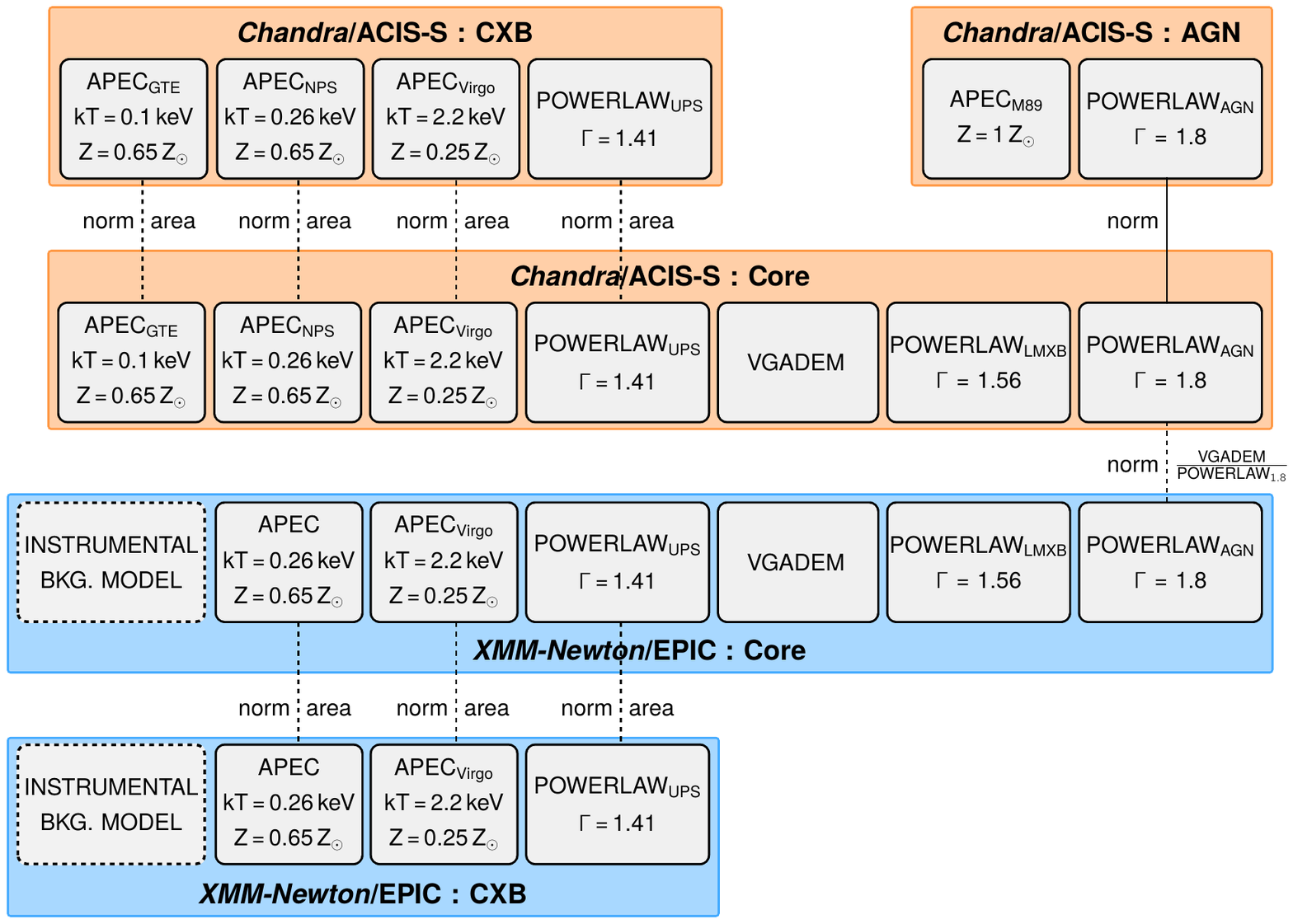}
    \caption{Modelling scheme of M89. The orange-coloured part represents the model for the \textit{Chandra} data extracted from the CXB, AGN and Core regions; while the blue-coloured part represents the \textit{XMM-Newton} model. The shared model components of the CXB and Core regions are tied with a scaling parameter obtained as a fraction of the areas of the two regions. A similar scaling parameter between the AGN \code{powerlaw} normalisation and the core \code{vgadem} normalisation was applied from the obtained ratio from the \textit{Chandra} fit into the \textit{XMM-Newton} model. All the other parameters for \textit{Chandra} and \textit{XMM-Newton} data are fitted separately.}
    \label{fig:fitting_scheme}
\end{figure*}

\section{Spectral analysis}\label{sec:sec_3_spectral_analysis}

Modelling of \textit{Chandra} and \textit{XMM-Newton} data is done separately. However, data from individual \textit{Chandra} observations as well as individual \textit{XMM-Newton}-EPIC detector (MOS1, MOS2 and pn) data are fitted simultaneously.
Both EPIC and ACIS-S data are fitted within the $0.55-7$ keV energy range. The 0.55 keV limit is chosen to avoid biases caused by the oxygen K-edge. In the case of RGS data, we chose the interval $8-27$ \AA. We used optimal binning by \citet{Kaastra_2016} for ACIS-S, EPIC, and RGS data.

The spectral modelling was performed using the Levenberg-Marquardt algorithm and Cash statistics \citep{Cash1979} within \code{PyXspec v2.1.0} \citep[Xspec 12.12.1;][]{Arnaud1996}. After finding the best-fit values, parameter uncertainties were estimated from posterior distributions obtained from an MCMC simulation. 

\subsection{Emission from galaxy core and tail}
\label{section:core}

There are three main sources of X-ray emission that originate inside M89. Therefore the following components are present in the Core and Tail regions exclusively: the first is the hot atmosphere that we are probing, which is assumed to be a thermal plasma in collisional ionisation equilibrium (CIE). The other two are the AGN component and the emission from Low-Mass X-ray Binaries (LMXBs).

The AGN component is represented by a \code{powerlaw} model with a photon index $\Gamma$ of $\sim 1.8$, derived from the \textit{Chandra}/ACIS-S spectrum (further described in Section \ref{AGN}). The integrated LMXB emission is reproduced by a \code{powerlaw} with a photon index fixed to 1.56 \citep{Irwin_2003, Su_2017}. As for the hot atmosphere, we tested three different models to represent the thermal emission component: single temperature (1T vapec), double temperature (2T vapec), and finally, multi-temperature, using the Gaussian distribution of emission measure (vgadem).
All the above-mentioned emission components are absorbed by the interstellar medium of our Galaxy, which is represented by the \code{phabs} model with a fixed value of $N_{H} \approx 1.36 \times 10^{20} \: \text{cm}^{-2}$ \citep{HI4PI2016} and our model is corrected for the redshift value of $z = 0.000113$, taken from the \textit{Nasa Extragalactic Database}. The total model representing the X-ray emission from the Core and Tail regions is as follows
\begin{center}
\begin{math}
\code{phabs * (vgadem + }\code{powerlaw}_{\code{AGN}} \code{+ powerlaw}_{\code{LMXB}}\code{)}.
    \end{math}
\end{center}

\subsubsection{vgadem}

As discussed in previous studies (see e.g. \citealp{Mernier_2015}, \citealp{Werner_2006}), a model employing the Gaussian distribution of temperatures improves statistics significantly for cluster cores. For the galaxy core, especially for our region that encircles a relatively wide area with presumably different temperatures, we employ the same model. In this analysis, we also tried single-temperature (1T) and double-temperature (2T) models to represent the thermal emission, and the results are presented in Table~\ref{table:parameters}. 

\subsection{Background modelling for EPIC and ACIS-S}
\label{section:background}

Unlike some of the previous abundance measurements of M89 \citep{Ji2009}, which subtracted background spectra, we carefully applied customised modelling for all of the background components. This approach has been shown to provide more reliable measurements compared to subtracting a local background spectrum (see relevant discussions in e.g. \cite{Mernier2015, Zhang2020, Zhang_2018}). In particular, \cite{dePlaa2007} reported that any small inaccuracies in the background spectra, which are highly probable when extracting spectra from a rigid background, can lead to inaccuracies in the temperature profile of an extended source. Therefore, because line emissivities depend on the temperature of the plasma at fixed abundances, the assumed temperature significantly affects the abundance determination. To avoid such inaccuracies in our abundance measurements, we chose to model the background components for both EPIC and ACIS-S.
 
Background effects that are not galaxy-originated are modelled by linking the parameters of the region of interest (Core or Tail region) and CXB region for the shared background emission components. 
The background components that are shared with the region of interest and CXB regions are the Virgo ICM, Unresolved Point Sources (UPS), Galactic Thermal Emission (GTE) and North Polar Spur (NPS) emissions. NPS is a structure in the Milky Way that emits both X-ray and radio emission, which is argued to be originated from a nearby supernova explosion (e.g. \citealp{Berkhuijsen_1971}), or that it is a remnant of an explosion or a starburst at the galactic centre (e.g. \citealp{Sofue_1994}). It is located at the northeastern edge of the Galactic bubble, with the brightest ridge at $(l, b) \sim (30^\circ, 20^\circ)$. As the Virgo cluster is in the vicinity of the NPS, we expect our observation to be affected by this emission. To model the background emission present in both the Core and CXB regions, we used the following model:
\begin{center}
\begin{math}
\code{phabs\,*\,}(\code{apec}_{\code{NPS}})\;+\;\code{phabs\,*\,}(\code{apec}_{\code{Virgo}}\code{+ powerlaw}_{\code{UPS}}).
    \end{math}
\end{center}
The NPS was modelled with an absorbed \code{apec} component ($kT_{\text{NPS}} = 0.26\,$keV, $Z_{\text{NPS}} = 0.65$\,Solar;  \citealp{Willingale2003}), where the absorption hydrogen density column was set to the half of the value measured by \cite{HI4PI2016} because the NPS emission is expected to lie behind at least 50 per cent of the line-of-sight Galactic column density \citep{Willingale2003}. The ICM emission from the Virgo cluster is represented by an absorbed \code{apec} component ($kT_{\text{Virgo}} = 2.2\,$keV, $Z_{\text{Virgo}} = 0.25$\,Solar) taken from \cite{Machacek_2006a} and checked independently for the CXB region by a separate ICM emission model. To account for the hard X-ray emission produced by UPS, we used a \code{powerlaw} component with photon index frozen to $\Gamma = 1.41$ \citep{deLuca_2004}, which represents well the contribution of background AGNs. The other two conventionally modelled background components of Local Hot Bubble (LHB) and the Galactic Thermal Emission (GTE) were also added, however, LHB could not have been constrained in both EPIC and ACIS-S, while the GTE could have been constrained only in the ACIS-S data. This is expected as we only included data above 0.55 keV, the contribution of LHB and GTE is expected to be sub-dominant, and the NPS and Virgo ICM are significantly dominant in M89 spectra. Moreover, the \code{apec} model ($kT_{\text{NPS}} = 0.26\,$keV) that reproduces the NPS is expected to represent the LHB and GTE emission components which have a similar temperature structure. Soft GTE emission in ACIS-S data is modelled with an \code{apec} component with the same abundance value and absorption hydrogen density column with the NPS component ($kT_{\text{GTE}} = 0.1\,$keV, $Z_{\text{GTE}} = 0.65$\,Solar).

 For testing purposes, we also tried removing the NPS component and instead having LHB and GTE components solely. However, this representation gave a significantly poorer fit. 

Spectra from the regions of interest (Core and Tail regions) and the CXB region were fitted simultaneously, and normalizations of all of their common components (NPS, Virgo and UPS) were tied by a scaling factor compensating the differences in effective areas between those regions, assuming that the Virgo ICM and NPS emission do not vary between the two regions. This approach was applied separately to \textit{XMM-Newton}/EPIC and \textit{Chandra}/ACIS-S data. 



For \textit{XMM-Newton}-EPIC data, there are also non-X-ray background components which should be modelled. 
HP background creates fluorescence emission lines in the spectra, even when the filter wheel is closed. The instrumental HP background is represented by a series of Gaussian lines and one broken powerlaw, unconvolved by the auxiliary response file, with values obtained by fitting the filter wheel closed (FWC) data, taken from Breuer et al. (in preparation). Because the HP background varies differently across the instrument, its modelling is done for the Core and CXB regions separately with different normalization values. 

The instrumental background model is further combined with the source model using a constant factor, which is allowed to vary during the fitting\footnote{We also tried unfreezing the normalization of individual instrumental lines to decrease the residuals, however, consistent results were obtained.}. Instrumental background modelling is done separately for the Core and CXB region spectra.

\subsection{AGN treatment approach}\label{AGN}

Due to the spatial resolution of the \textit{XMM-Newton} observation, we were not able to detect and remove the central AGN region from our spectra for the \textit{XMM-Newton}-EPIC data. Therefore, we were required to reproduce the AGN emission in our model. Thanks to the excellent spatial resolution of the \textit{Chandra} X-ray observatory, we were able to extract a 1.5 arcsec circular region covering the AGN from the \textit{Chandra}/ACIS-S data. By fitting the spectra of this region alone, we estimated the photon index of the \code{powerlaw} model that is to reproduce the central AGN emission. We use this model in the \textit{XMM-Newton}/EPIC and \textit{Chandra}/ACIS-S fittings for the Core region.

\begin{figure}
    \centering
    \includegraphics[width=\columnwidth]{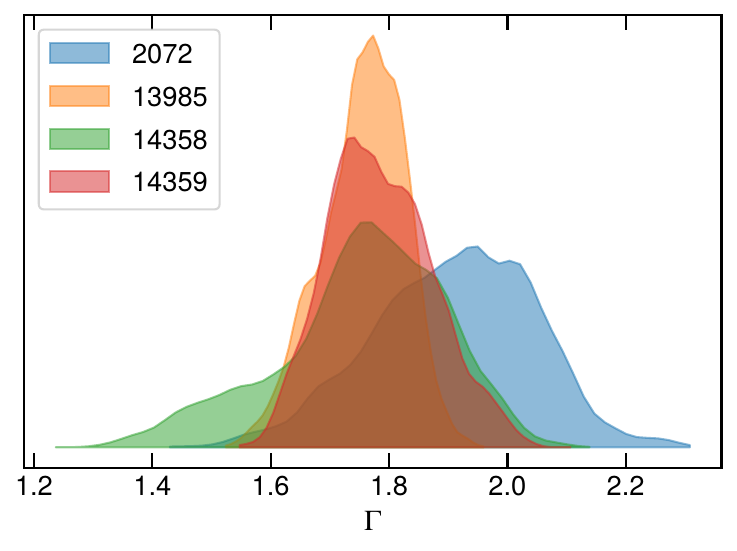}
    \caption{Posterior distributions of the powerlaw photon index $\Gamma$ for the central 1.5 arcsec region (AGN region) for individual \textit{Chandra} OBSIDs. All distributions are consistent within uncertainties and therefore were combined resulting in the final value of $\Gamma = 1.79^{+0.13}_{-0.11}$.}
    \label{fig:phoindex}
\end{figure}

Using the central AGN region (1.5 arcsec), we investigated the emission from the central AGN and its variability within four individual \textit{Chandra} observations. All observations were simultaneously fitted with an absorbed \code{powerlaw} model with an additional thermal component, \code{apec}, representing the CIE emission from this region:
\begin{center}
    \code{phabs * (powerlaw + apec)}.
\end{center}
The \code{apec} abundance was fixed to Solar and the temperatures and normalizations were for all observations tied with respect to the first one. All other parameters (\code{powerlaw} photon index and normalization) were untied within observations and left free during the fitting.

Photon indices of the \code{powerlaw} component for individual \textit{Chandra} observations were consistent within uncertainties (Figure \ref{fig:phoindex}) so we stacked their posterior distributions resulting in a final value of $1.79^{+0.13}_{-0.11}$. The normalizations of the \code{powerlaw} component varied slightly within individual observations, but the differences between individual OBSIDs were of a factor of $\sim 2$ at most. The total unabsorbed powerlaw flux averaged over individual OBSIDs was estimated to $F_{2-10 \: \text{keV}} = (1.7 \pm 0.1) \times 10^{-14} \: \text{erg} \, \text{cm}^{-2} \, \text{s}^{-1}$. Assuming the distance of 15.9 Mpc \citep{Tully2013}, this corresponds to a resulting AGN luminosity of $L_{2-10 \: \text{keV}} = (5.5 \pm 0.3) \times 10^{38} \: \text{erg s}^{-1}$.

\begin{figure}
    \centering
    \includegraphics[width=\columnwidth]{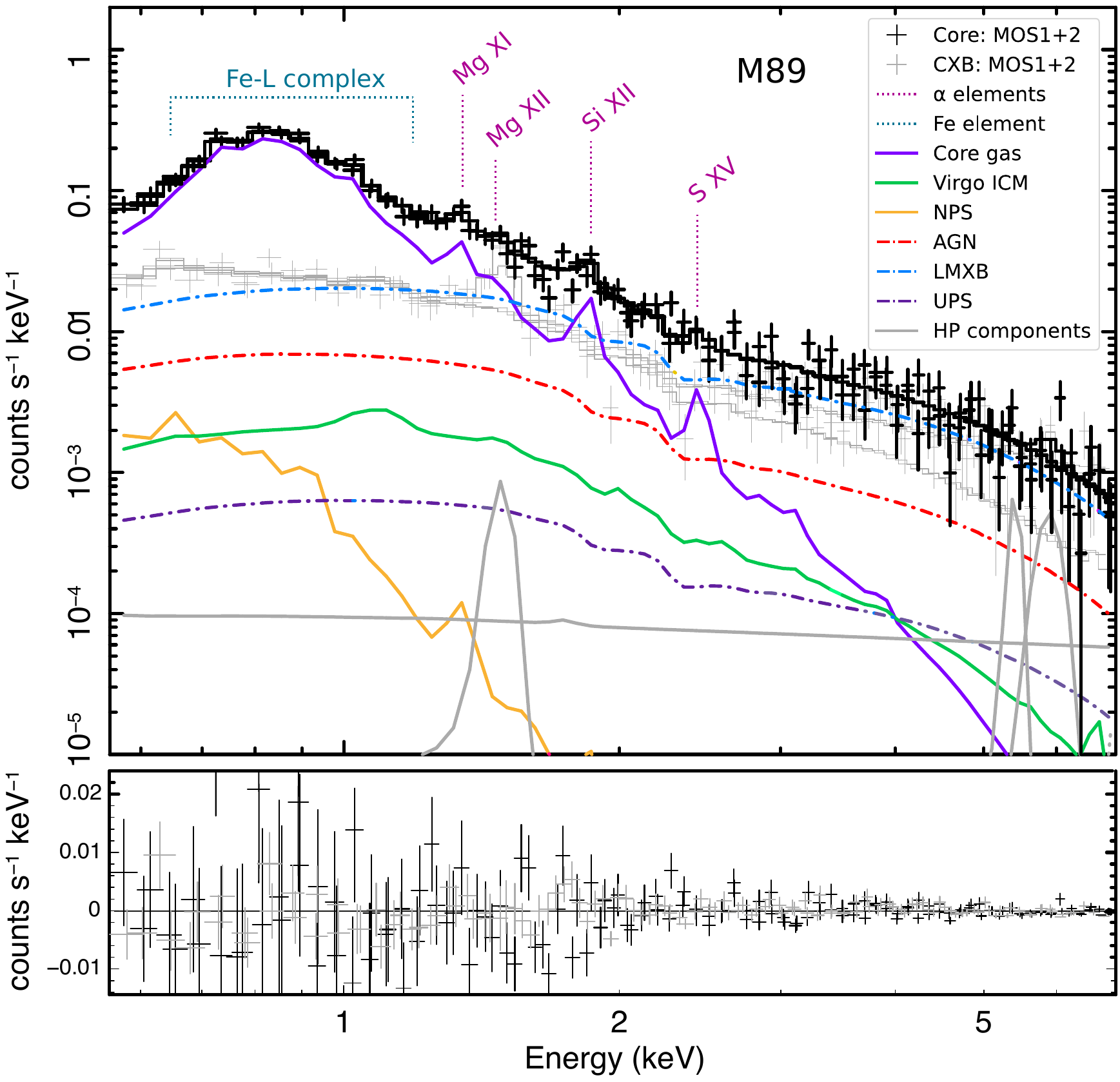}
    \caption{\textit{XMM-Newton}-EPIC MOS 1 and MOS 2 combined fit of the core of M89. The CXB region data, which is fitted simultaneously with the Core by tying the parameters of common components after scaling, is also shown (light grey). The best-fit models of the M89 core gas (purple), Virgo ICM (green), NPS (yellow), AGN (dashed-red), LMXB (dashed-blue), UPS (dashed-dark purple) and HP (for MOS 1 and MOS 2) components (grey) are also presented.}
    \label{fig:modelplot}
\end{figure}

As the central AGN emission can not be spatially resolved and excluded with \textit{XMM-Newton}, we instead constrain the contribution of the central AGN emission. For \textit{Chandra} data, we fitted the central Core region (in this case, including the central point source) simultaneously with the AGN region, and we tied the normalizations of their \code{powerlaw} components, assuming that all emission of the AGN lies within the central 1.5 arcsec region\footnote{For the core region, we assumed similar model as in Section \ref{section:core}.}. For the Core region, we then expressed the \code{powerlaw} normalization as a constant ratio with respect to the normalization of the main thermal emission component (\code{vgadem}), which was allowed to vary. 
This ratio was then assumed to be the same for \textit{XMM-Newton} data, and the uncertainty was 
propagated by fitting the \textit{Chandra} spectra simultaneously with that of \textit{XMM-Newton}. We note that, in general, the contribution of the AGN can be time-variable and the ratio of thermal to AGN components can vary from observation to observation. Still, we believe that this represents a better approach than fitting the AGN \code{powerlaw} component from \textit{XMM-Newton} data alone. For comparison, we performed the spectral fitting for \textit{XMM-Newton} data also without the AGN \code{powerlaw} component. In this case, statistical uncertainties were larger. In Figure~\ref{fig:modelplot}, we present MOS spectra of the Core and CXB regions with all emission components. 

\subsection{RGS Analysis}

The RGS data is expected to contain very limited background contamination after the standard background template derived from CCD9 was subtracted, which registers the least source events. Therefore, we fitted RGS data without any background components. 


We tried several fitting combinations of 1) including the second order spectra or restricting to the first order exclusively; 2) testing 1T, 2T, Gaussian temperature distribution (\code{vgadem}). The complete results are presented in Table \ref{table:parameters} and further demonstrated in Appendix~\ref{appendix:a}.


\section{Results}
\label{sec:Results}

Abundance ratios in the Core region were obtained by separate modelling of {\it XMM-Newton}/EPIC, {\it XMM-Newton}/RGS and {\it Chandra}/ACIS-S spectra. In addition to the iron (Fe), the magnesium (Mg), silicon (Si) and sulphur (S) abundances were allowed to be free in the EPIC CCDs. The neon (Ne) abundance was initially left free to vary for both the EPIC and ACIS-S models, but it could only be constrained using ACIS-S data, so ultimately neon was set free in the ACIS-S model and fixed relative to iron (Fe) in the EPIC model. For RGS, the oxygen (O), neon (Ne), magnesium (Mg) and iron (Fe) were the free abundance parameters. First, we see that fitting with the addition of second-order RGS spectra gives better statistics, also making the results from the different temperature models more consistent with each other. In each of these fits, all the other elements were tied to Fe except He, which was fixed to be Solar.

After performing our multi-temperature modelling, we measured indications for possible super-Solar abundance ratios (i.e. $\alpha$/Fe > 1) of Ne/Fe, Mg/Fe, Si/Fe and S/Fe in ACIS-S and EPIC data. As for RGS measurements, Mg/Fe is also super-Solar, however, the O/Fe and Ne/Fe ratios are found to be sub-Solar and Solar, respectively, within uncertainties.

Additionally, we performed tests on the removal of point sources from the CXB region spectra of EPIC using PSF maps with sizes of 20 arcsec, approximately corresponding to a 0.75 ECF. We see that the elemental abundance and abundance ratio results remain consistent even with the application of larger point source masks.

We also tested the differences obtained from the blank-sky data set instead of stowed spectral files. After performing the spectral fit, we observed that the results are comparable, although uncertainties increase with the background-removed blank-sky spectra.

EPIC studies of M89 were previously conducted by \cite{Ji2009}, as also stated in Section~\ref{sec:introduction}. They utilized a single-temperature model, i.e. \code{wabs $\times$ (vapec + powerlaw)}, and subtracted a local background spectrum. Although they investigated a region with a similar size (1 arcmin), we observe that none of our individual abundances from multi-temperature models agree with their single-temperature results. This is expected, because of the `Fe bias' as explained in Section \ref{sec:introduction}. Similarly, although we found super-Solar Mg/Fe, Si/Fe, and S/Fe, they reported Solar ratios.

When we also conducted a single-temperature modelling, we observed that only our individual Mg and Si abundance measurements agree with their results, which were $0.46_{-0.10}^{+0.11}$ and $0.44_{-0.12}^{+0.13}$ Solar, respectively. We also note that the previous versions of AtomDB \citep{Smith2001, Foster2012} before the update of AtomDB 3.0.9 in 2020 suffered from the interpolation issue\footnote{\url{http://www.atomdb.org/interpolation/index.php}}. This issue involved discrepancies caused by the interpolation of \code{(v)apec} between its pre-calculated spectra \citep{Mernier2020}, which has now been fixed. Therefore, the possible reasons for the discrepancy, even in the single-temperature model, are (i) the use of outdated spectral codes and databases by \cite{Ji2009}, (ii) differences in the subtraction of background spectra, which could create biases in abundance measurements \citep[see][]{dePlaa2007}, and 
(iii) differences in modelling the AGN and LMXB components, which might not be well-described by the single \code{powerlaw} model in \cite{Ji2009}.

Moreover, their RGS analysis suggests individual abundances around $4 - 5$ Solar. These values are highly unusual considering the highest recurrently and robustly reported elemental abundances are around $\sim 2$ Solar (e.g. Centaurus Cluster; \citep{Matsushita2007, Sanders2016, Fukushima2022}).

\begin{figure}
    \centering
    \includegraphics[width=\columnwidth]{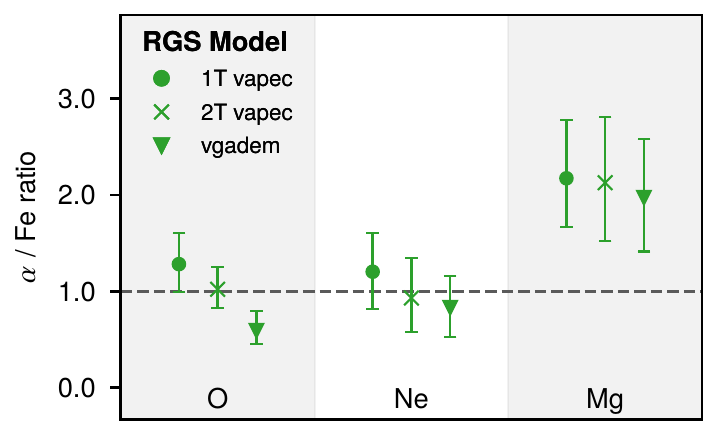}
    \caption{O/Fe bias in the \textit{XMM-Newton}-RGS data of the low temperature plasma. The O/Fe ratios differ significantly between multi-temperature models. }
    \label{fig:rgs_model}
\end{figure}

In the CHEERS catalogue, \cite{dePlaa__2017} showed that M89 has an O/Fe ratio of $2.0\pm0.6$, which is the highest O/Fe ratio observed in the whole sample. They also showed Ne/Fe = $1.1\pm0.7$, and subsequent work by \cite{Mernier2016a} indicates Si/Fe = $1.4_{-0.7}^{+5.3}$. The main difference between our sub-Solar O/Fe ratio and their super-Solar measurement is the choice of the model: we used a multi-temperature model with Gaussian distribution of emission measure while \citet{dePlaa2017} assumed single-temperature plasma. Therefore, we suspect that in the analysis of low-temperature M89 RGS spectra, the choice of modelling (single-T vs. multi-T) may have a considerable impact on the O/Fe ratio and its interpretation. In our results, we see that the O/Fe ratio changes $> 1\sigma$ between single temperature to \code{vgadem} model, which is presented in Figure~\ref{fig:rgs_model}. The differences between the CHEERS results and ours may also be attributed to the use of the AtomDB database and \code{vgadem}/\code{vapec} plasma codes in our study, whereas they employed the SPEXACT database and \code{cie} plasma code.


The complete list of parameters for EPIC, RGS and ACIS-S obtained from different multi-temperature models is presented in Table~\ref{table:parameters}. In the following sections, we first describe the fitting biases. Then, we present the chemical composition results within the Core and Tail regions.

\setlength{\tabcolsep}{5.5pt}
\renewcommand{\arraystretch}{1.65}
\begin{table*}
\caption{The parameters derived from different temperature models for \textit{XMM-Newton}/EPIC, \textit{XMM-Newton}/RGS and \textit{Chandra}/ACIS-S observations of M89. Elemental abundances are in the Solar units. The background modelling is the same for each temperature model.}
\begin{tabular}{lccc|ccc|ccc}
     &  \multicolumn{3}{c}{\textit{XMM-Newton} EPIC}      &    \multicolumn{3}{c}{\textit{XMM-Newton} RGS}    &  \multicolumn{3}{c}{\textit{Chandra} ACIS-S} \\ \hline 
Parameter    & vapec & 2 $\times$ vapec & vgadem & vapec & 2 x vapec & vgadem & vapec & 2 x vapec & vgadem \\ \hline
kT$_1$ (keV)      &   $0.71\pm0.01$      &  $0.62\pm0.01$         &  --      &   $0.61^{+0.02}_{-0.03}$     & $0.84\pm0.09$     &   --     &   $0.65 \pm 0.06$    &     $0.35 \pm 0.02$       &    --    \\
kT$_2$ (keV)  &  --  &  $0.83\pm0.03$  &   --     &   --    &    $0.53_{-0.03}^{+0.02}$  &  --   & --  &     $0.70 \pm 0.04$       &  --  \\
kT$_{\mu}$ (keV)  &    --   &    --      &    $0.64\pm0.02$      &  --    &   --        &  $0.49_{-0.10}^{+0.08}$      &   --    & --  & $0.49 \pm 0.03$    \\
kT$_{\sigma}$ (keV) &   --    &     --      &    $0.26\pm0.04$      &   --    &   --        &   $0.41_{-0.14}^{+0.19}$     &   --    & -- &   $0.33 \pm 0.04$     \\
O             &   --    &    --       &    --    & $0.29\pm0.08$       &   $0.29\pm0.08$        &   $0.21^{+0.06}_{-0.05}$      &   --    &     --      &   --    \\
Ne            & --      &     --      & --       &  $0.27^{+0.10}_{-0.09}$      &     $0.26\pm0.11$      &  $0.30^{+0.09}_{-0.11}$       &  $0.99_{-0.11}^{+0.09}$    &      $0.94_{-0.15}^{+0.25}$     &   $1.51_{-0.26}^{+0.30}$     \\
Mg            &  $0.67\pm0.14$      &  $1.21_{-0.27}^{+0.23}$  &  $0.71_{-0.15}^{+0.14}$  &  $0.50^{+0.15}_{-0.14} $  &  $0.60\pm0.19$  &  $0.69_{-0.12}^{+0.20}$ &  $0.64 \pm 0.07$  &  $0.99_{-0.11}^{+0.14}$  &  $1.58_{-0.25}^{+0.41}$   \\ \vspace{1mm}
Si            &  $0.55_{-0.12}^{+0.14}$       &   $1.07_{-0.19}^{+0.18}$         &    $0.73\pm0.13$      &  --     &  --         &   --     &   $0.46 \pm 0.05$  &  $0.73_{-0.10}^{+0.11}$  &  $1.18_{-0.22}^{+0.31}$  \\
S             &  $1.36\pm0.46$      &  $1.92_{-0.71}^{+0.79}$          &    $1.07\pm0.48$      &   --    &   --       &  --   &  $0.76_{-0.21}^{+0.25}$  &  $1.04_{-0.27}^{+0.30}$  &   $1.57_{-0.48}^{+0.70}$    \\
Fe            &  $0.62^{+0.08}_{-0.07}$      &    $1.05_{-0.14}^{+0.13}$        &   $0.58\pm0.07$      &  $0.23\pm0.04  $   &  $0.28\pm0.06$  &  $0.35_{-0.08}^{+0.09}$  &  $0.54 \pm 0.03$  &  $0.60 \pm 0.04$ &  $1.04_{-0.27}^{+0.30}$   \\
C-stat / d.o.f. & 437 / 382      & 420 / 380        &   414 / 381        &   649 / 534    &   642 / 533        &    638 / 533    &   759 / 513    &     659 / 511      &   672 / 512  \\ 
\hline
SNcc contribution & $81.4\%_{-5.8}^{+4.1}$ & $80.9\%_{-5.9}^{+4.2}$  & $85.6\%_{-4.4}^{+3.3}$  & $84.3\%_{-5.2}^{+3.7}$  & $79.9\%_{-6.7}^{+4.7}$  & $71.3\%_{-7.3}^{+5.3}$ & $82.4\%_{-2.0}^{+1.8}$  & $85.4\%_{-2.7}^{+2.2}$  & $85.7\%_{-5.3}^{+3.7}$    \\
$\rchi^2$ / d.o.f. & 4.9 / 2 & 3.3 / 2  & 0.7 / 2 & 2.7 / 2  & 2.8 / 2  & 4.6 / 2 & 24.0 / 3  & 1.7 / 3  & 1.1 / 3    \\
\hline
\end{tabular}
\label{table:parameters}
\end{table*}

\begin{figure*}
    \centering
    \includegraphics[width=0.3\textwidth]{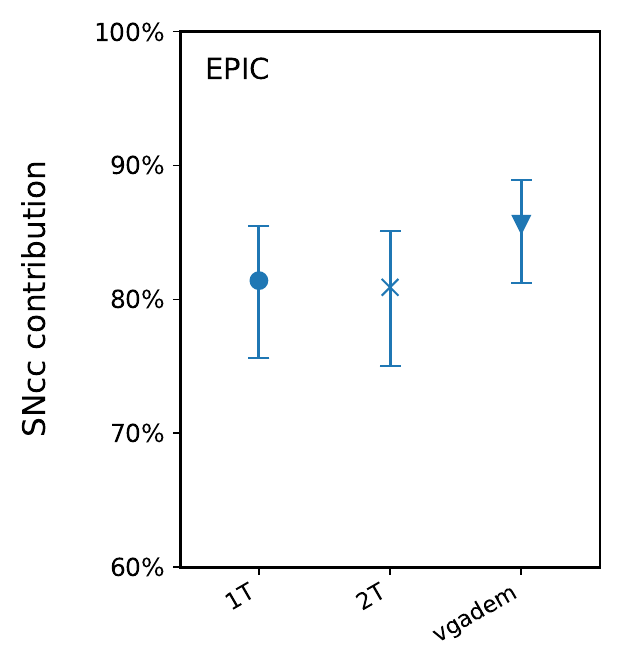} \hfill
    \includegraphics[scale=0.47]{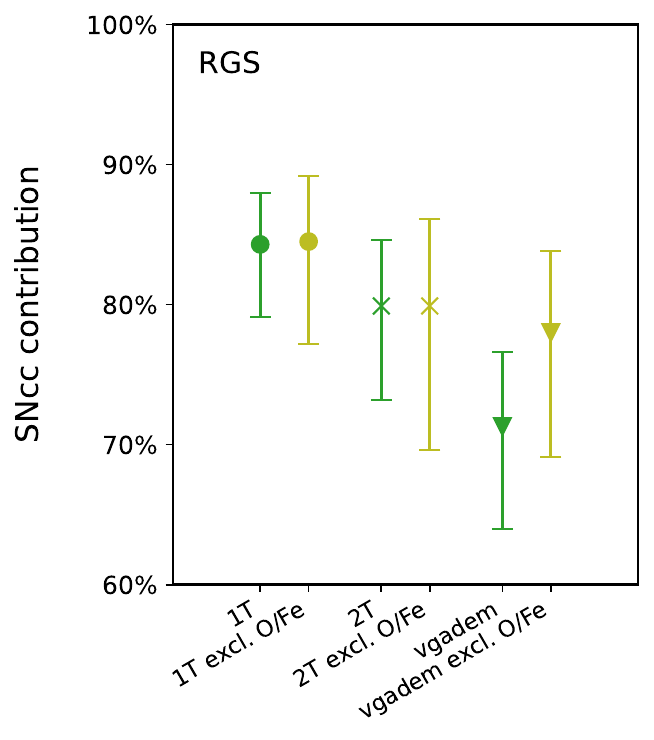}
    \hfill
    \includegraphics[width=0.3\textwidth]{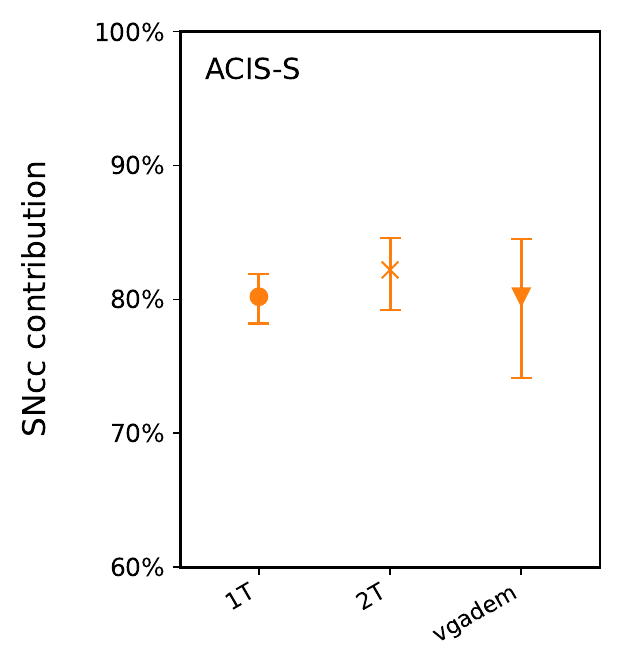}
    \caption{SNcc contribution of each instrument with different temperature models. (\textit{Left}) The EPIC calculation used Mg/Fe, Si/Fe and S/Fe ratios. (\textit{Middle}) The RGS calculation used O/Fe, Ne/Fe and Mg/Fe ratios. (\textit{Right}) The ACIS-S calculation used Ne/Fe, Mg/Fe, Si/Fe and S/Fe ratios.}
    \label{fig:SNcc}
\end{figure*}

\subsection{Systematic uncertainties}

In order to investigate the differences in abundance estimates between single and multi-temperature models, we refit our data from all three instruments with the same modelling scheme, changing only the M89 hot gas emission component: the single-temperature (\code{vapec}), double-temperature (\code{vapec+vapec}) and multi-temperature with Gaussian distribution of emission measure (\code{vgadem}). The resulting temperatures and abundances derived using individual models (1T, 2T, vgadem) and instruments (ACIS-S, EPIC, RGS) are presented in Table \ref{table:parameters} and the abundance estimates are graphically compared in Figure~\ref{fig:models_detectors}.

\subsubsection{O/Fe bias in the RGS data of low-temperature plasma}
\label{subsection:differences_btw_models}

The most prominent discrepancy is that the O/Fe ratio is $\sim2$ times lower in \code{vgadem} compared to 1T \code{vapec} model, as presented in Figure~\ref{fig:rgs_model}. 
Although not formally significant in our case, we note that this decreasing trend between 1T \code{vapec} and \code{vgadem} is also present in Ne/Fe and not in Mg/Fe ratios. 

We caution the reader that in the \code{vgadem} model, $kT_{\mu}$ and $\sigma_{kT}$ values are $0.49_{-0.10}^{+0.08}$ keV and $0.41_{-0.19}^{+0.14}$ keV respectively. As the dispersion of the Gaussian distribution is comparable to the mean, the imperfections in constraining the multi-temperature nature may cause additional measurement bias. 


As we have different abundance results for each instrument and model, we further check the SNe ratios derived from these values. Our aim is to investigate the bias more extensively.




\subsubsection{The relative SNcc contribution} 


In this section, we calculate the relative supernova contributions to the enrichment via the $\alpha$/Fe ratios. Considering various approaches to model M89's hot atmosphere, we fit the abundance ratios obtained from each instrument to a number of SNcc and SNIa nucleosynthesis yield models available from the literature. Our strategy is to compare the obtained best-fitting SNcc yields, checking for any discrepancy between different temperature models. We note that, for the calculation of SNcc fractions, the input elements and their abundance values differ for each instrument. Therefore, when comparing the relative contributions of SNcc between models, it is important to restrict the comparison within each instrument separately.

We aim to recover the total SNcc/(SNcc + SNIa) fraction contributing to the enrichment. For the calculation, we used \code{SNeratio}\footnote{\url{https://github.com/kiyami/sneratio}} code developed by \cite{Erdim_2021}, used by recent enrichment studies (e.g. \citealp{Gatuzz2023}). The \code{SNeratio} code provides the relative SNe contribution for the observed abundance data for selected yield tables and models. 

For SNIa yields, we applied 3D models of near Chandrasekhar-mass with delayed detonation from \cite{Seitenzahl_2012} and a pure deflagration model from \cite{Fink_2014}. For the SNcc yields, we used a model from \cite{Nomoto_2013}. For the yield calculation of the latter, we assume a Salpeter initial mass function (IMF) \citep{Salpeter_1955}. The lower zero-age main sequence mass of stars is adopted with $10 M_{\odot}$ and $50 M_{\odot}$ as the lower and upper limit, respectively. We integrate the yields using the equation,
\begin{equation}
M_i = \frac{\int_{10 M_{\odot}}^{50 M_{\odot}}M_i(m)m^{\alpha}dm}{\int_{10 M_{\odot}}^{50 M_{\odot}}m^{\alpha}dm},
\end{equation}
where $M_i$ is the total yield (also denoted as $Y_i$) of the $i$-th element, $M_i(m)$ is the $i$-th element mass produced in a star of mass $m$, and $\alpha$ is the slope of the IMF. For the parameter $\alpha$, we adopted the value of $-2.35$ corresponding to the Salpeter IMF.

Furthermore, the number of atoms of element X is given by
\begin{equation}
N_X = aN_{X,\text{SNIa}} + bN_{X,\text{SNcc}},
\end{equation}
where $N_{X,\text{SNIa}}$ and $N_{X,\text{SNcc}}$ are the number of atoms produced in a single SNIa and SNcc event, respectively, and $a$ and $b$ are the multiplicative factors. 

        
When calculating the SNe fraction with the \code{SNeratio} code, we used different initial metallicity values for the SNcc yields of Z = 0.0, 0.0001, 0.004, 0.008, 0.02, 0.05. The results are presented in Table~\ref{table:parameters} and Figure \ref{fig:SNcc}.

For EPIC calculations, we used the Mg/Fe, Si/Fe and S/Fe ratios and found that the SNcc contribution results from different temperature models agree with each other. In this sample, the \code{vgadem} result is the largest ($85.6^{+3.3}_{-4.4}\,\%$), while the 1T \code{vapec} is the smallest ($81.4^{+4.1}_{-5.8}\,\%$).

\begin{figure*}
\centering
    \includegraphics[width=0.76\textwidth]{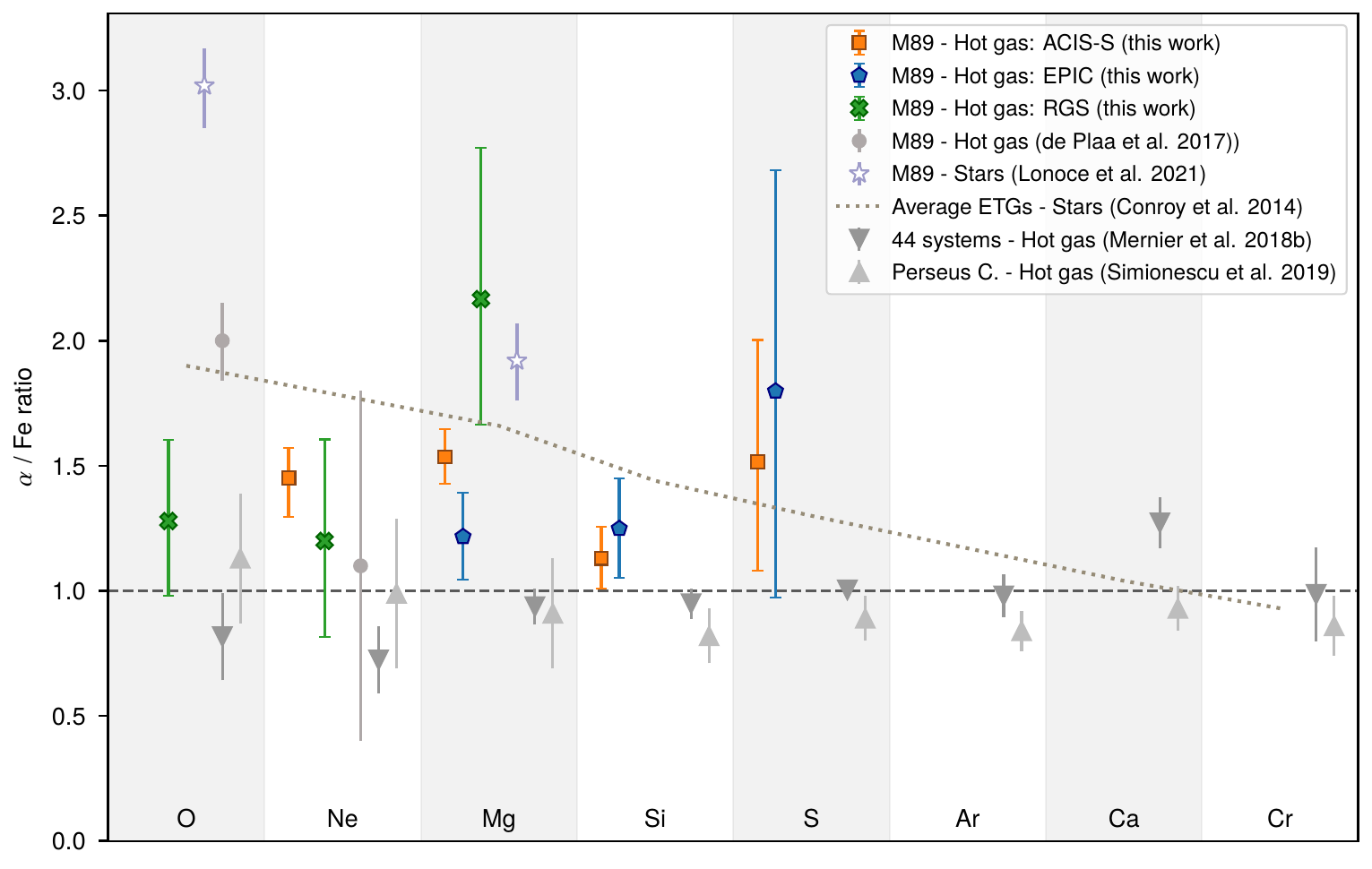}
    \caption{$\alpha$/Fe ratios in the core of the galaxy. For ACIS-S (orange); Ne/Fe, Mg/Fe, Si/Fe and S/Fe ratios are derived via \code{vgadem} model. For EPIC (blue), \code{vgadem} model is used to derive Mg/Fe, Si/Fe and S/Fe ratios. In the RGS results (green); O/Fe, Ne/Fe and Mg/Fe are measured with a 1T \code{vapec} model. }
    \label{fig:core_comparison}
\end{figure*}

As for the SNe contributions estimated from RGS data, we used O/Fe, Ne/Fe and Mg/Fe ratios. From these ratios, the obtained results for 1T \code{vapec} and 2T \code{vapec} are comparable within uncertainties. However, we observe > 1$\sigma$ discrepancy between \code{vgadem} and 1T \code{vapec} results for SNcc contributions. We also performed the calculation by excluding the biased O/Fe ratio as input. In this case, we observe that all SNcc contributions agree within uncertainties. We also note that, although the results are consistent when excluding the O/Fe ratios, the same aforementioned trend persists, where the 1T \code{vapec} and \code{vgadem} are, respectively, the largest and smallest.

The ACIS-S calculation used Ne/Fe, Mg/Fe, Si/Fe and S/Fe ratios. Using these ratios, we again derived comparable SNcc contributions between temperature models. In this case, the 2T \code{vapec + vapec} model gave the highest percentage ($85.4^{+2.2}_{-2.7}\,\%$). Nevertheless, we found that SNcc contribution is $> 70\%$ for all detectors and temperature models. 







\subsubsection{Summary on systematic uncertainties}

The discussion on systematic uncertainties can be summarised as follows:

\begin{itemize}
    \item The O/Fe ratio changes significantly (> 1$\sigma$) in the RGS data 
    between different multi-temperature models.
    \item We observe a systematic change in the SNcc ratios, with or without the O/Fe ratio.
    \item The SNcc ratios obtained from RGS data differ by more than 1$\sigma$ between different temperature models. However, excluding the O/Fe ratio makes the results comparable. 
    \item The SNcc contribution cannot be well constrained with the accessible abundance ratios. 
    \item Due to the relatively low clean exposure time of \textit{XMM-Newton}, we observe rather high statistical uncertainties that mask out the true nature of the systematic uncertainties. 
    \item With deeper observations and consequently less statistical error, it would be possible to understand the systematic differences between temperature models more accurately. However, measuring abundance ratios and constraining SNe ratios robustly would still be difficult even if we have deeper observations with the current spectral resolution available.   
\end{itemize}

Consequently, because of the systematic biases and the fact that the $\sigma_{kT}$ of the \code{vgadem} model ($0.41_{-0.19}^{+0.14}$ keV) is comparable to the $kT_{\mu}$ value ($0.49_{-0.10}^{+0.08}$ keV), we conclude that, for RGS, the \code{vgadem} model is not adequate to constrain the temperature structure. Therefore, for further analysis in this paper, we use the 1T model for the RGS data, which is also used in \cite{dePlaa2017} for M89. 

\subsection{Summary on $\alpha$/Fe ratios}

\subsubsection{Core}

In Figure~\ref{fig:core_comparison}, we present the best-fit values of the abundance ratios that are previously discussed. In the plot, the results for EPIC and ACIS data correspond to \code{vgadem} model, while for RGS a single-temperature (1T vapec) fit was presented.

We find the overall abundance ratios to be super-Solar. S/Fe and Si/Fe ratios from ACIS-S and EPIC are $1\sigma$ larger than the Solar values. Similarly, Mg/Fe for EPIC and RGS are also $1\sigma$ larger while the ACIS ratio of Mg/Fe is $5\sigma$ greater. The Ne/Fe ratio is again super-Solar for both ACIS-S and RGS. Likewise, the O/Fe ratio is $1\sigma$ higher than its Solar value in RGS.

The super-Solar O/Fe and Ne/Fe ratios, also presented in Figure~\ref{fig:core_comparison}, agree with the results of \cite{dePlaa2017}. Moreover, \cite{Mernier2016a} showed that Si/Fe is also super-Solar with a value of $1.4^{+5.3}_{-0.7}$; however, due to a large positive uncertainty, we exclude it from the plot for visualisation purposes. In fact, M89 is the only sample in the CHEERS measurements having super-Solar values for all ratios.

This result differs from the global Solar composition in the elliptical galaxy atmospheres \citep{Mernier2018c}. The results indicate that the composition in M89 hot gas is more similar to the stellar components. 

The physical interpretation of these results and the possible scenarios explaining the atmosphere with super-Solar abundance ratios are presented and discussed in the Discussion section.

\subsubsection{Tail}

\begin{figure}
    \centering
    \includegraphics[width=\columnwidth]{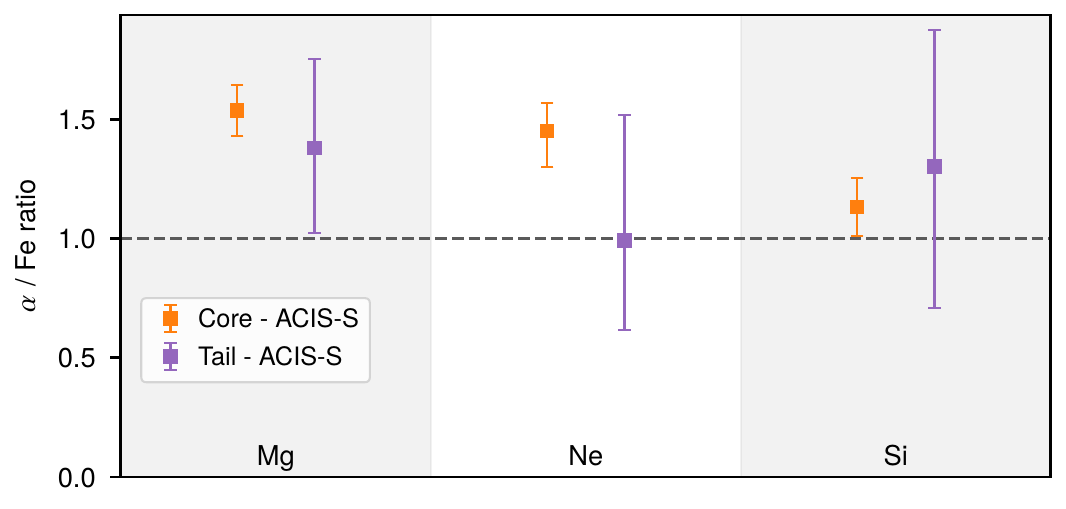}
    \caption {$\alpha$/Fe ratios in the Tail region obtained from the \textit{Chandra}/ACIS-S data.}
    \label{fig:tail_ratios}
\end{figure}

For the striped tail of the galaxy, the chemical abundances could only be constrained with the deep \textit{Chandra}/ACIS-S observation. We found that absolute abundances are lower than in the Core region and thus less than Solar. The ratios, on the other hand, are super-Solar for Mg/Fe and Si/Fe, all comparable with the Core region within uncertainties, presented in Figure~\ref{fig:tail_ratios}. Ne/Fe ratio, on the other hand, is Solar but still comparable with that of the Core region. Regrettably, the uncertainties in the tail region are so large that a robust physical interpretation is not possible.

The chemical abundance ratios in the two gas regions are found to be comparable. These results are further discussed in the Discussion section.

\subsection{AGN activity}

\subsubsection{The AGN driven shock front}

\begin{figure}
    \centering
    \includegraphics[width=\columnwidth]{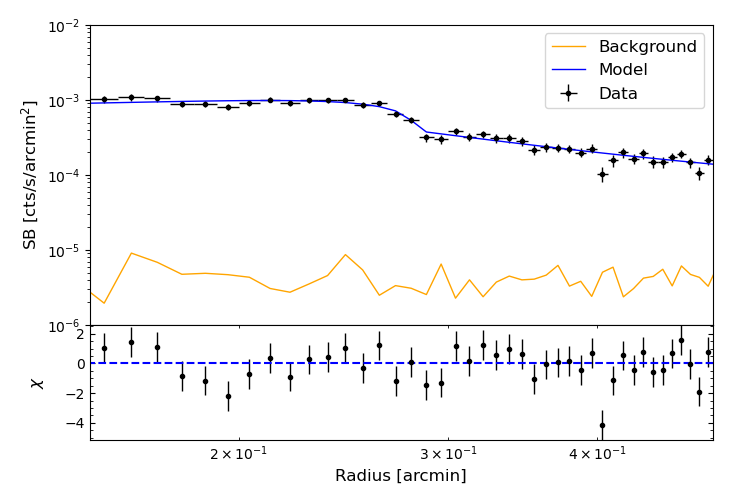}
    \includegraphics[width=\columnwidth]{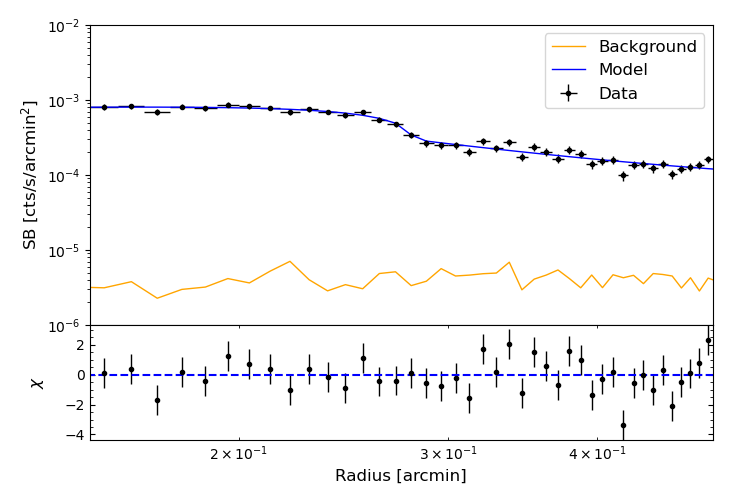}
    \caption{\textit{(Top)} The best fit broken powerlaw model fitted to the surface brightness profile of the northern ring, extracted from an annular section spanning from $55^{\circ}$ to $145^{\circ}$ with inner and outer radii of 11 and 31 arcsec. \textit{(Bottom)} The best fit broken powerlaw model fitted to the surface brightness profile of the southern ring, extracted from an annular section spanning from $245^{\circ}$ to $335^{\circ}$ with inner and outer radii of 11 and 31 arcsec.}
    \label{fig:sbp}
\end{figure}

One of the prominent features of M89 is an `hourglass' structure first noted by \cite{Filho2004} in the inner region of the galaxy. This hourglass shape is composed of two rings of approximately circular shape where each inner edge reaches $\sim0.85~\mathrm{kpc}$ while the outer edge reaches $\sim1.30~\mathrm{kpc}$ from the centre. These rings of shocked gas have been attributed by past research to the nuclear activity of the central AGN of this galaxy \citep{Machacek_2006b}.


To determine the temperature jump across the rim of the shock, we utilised all four \textit{Chandra} observations with a total observation time of 201.3 ks, in contrast to \cite{Machacek_2006b} who used a single 54.4 ks observation. Unlike other studies, we investigate the Northern and Southern Rings separately for the first time. We fit the spectra of the regions NR, NRO, SR, and SRO described in Section~\ref{region_selection} with a single-temperature (1T \code{vapec}) model with abundances fixed to Solar values, focusing only on the hot atmosphere component as described in Section~\ref{sec:sec_3_spectral_analysis}. We also tried spectral fitting using the single temperature Fe abundance result obtained by 1T \code{vapec} fitting of the \textit{Chandra} Core region data, which is approximately half of the Solar value. However, we see that changing the abundance value affects the temperature value by less than 0.5\%, and we ultimately decided to use the Solar abundances. The resulting temperatures are $0.73\pm0.01$ keV at the rim of the northern ring (NR), with the temperature decreasing to $0.43\pm0.02$ keV on the outside of the northern ring (NRO). For the southern part, the temperature reaches $0.76\pm 0.01$ keV at the edge of the shock (SR) and declines to $0.43\pm0.03$ keV on the outside of the rim (SRO).

The change in electron density across the edge of the shock can be gauged from the surface brightness profile across these rings. Therefore, we extract the surface brightness profile from the exposure-corrected, background-subtracted image in the energy band of $0.5-2.0$ keV and analyse each ring in annular sections encompassing both the NR (SR) and NRO (SRO) regions. For each brightness discontinuity, we model the surface brightness profile using an MCMC analysis and we assume a spherically symmetric broken powerlaw distribution projected along the line of sight for the electron density of  
\begin{equation}\label{eqn:eDens}
    n_e =
    \begin{cases}
    n_{\mathrm{rim}}\left(\frac{r}{r_{\mathrm{rim}}}\right)^{-\alpha_1}   & r < r_{\mathrm{rim}}\\
    \frac{1}{J}n_{\mathrm{out}}\left(\frac{r}{r_{\mathrm{out}}}\right)^{-\alpha_2}  & r \geq r_{\mathrm{rim}}\mathrm{~,}
    \end{cases}
\end{equation}
where $r_{\mathrm{rim}}$ is the outer radius of the shock ring, $r_{\mathrm{out}} = 3.1$ kpc and marks the unshocked region outside of the rim, $n_{\mathrm{rim}}$, $n_{\mathrm{out}}$ are the normalisation inside and outside the rim of the shock, $\alpha_1$, $\alpha_2$ are the power law indices for each region, while $J$ is the discontinuity parameter. The discontinuity parameter is given by
\begin{equation}
    J = \frac{\Lambda_{\mathrm{rim}}n^2_{\mathrm{rim}}}{\Lambda_{\mathrm{out}}n^2_{\mathrm{out}}}\mathrm{~,}
\end{equation}
where $\Lambda$ and $n$ represent the emissivity and electron density of each region.

We find the best-fit position for the outer rim of the shock to be $r_{\mathrm{rim}} = 1.31_{-0.01}^{+0.01}$ kpc for the northern ring and $r_{\mathrm{rim}} = 1.29_{-0.02}^{+0.01}$ kpc for the southern ring. The inferred electron density ratio $n_{\mathrm{rim}}/n_{\mathrm{out}}$ across each ring is $ 1.46_{-0.09}^{+0.09}$ (NR) and $1.41_{-0.08}^{+0.09}$ (SR). By fitting the region inside and outside of the shock rim with a simple powerlaw, we calculate the densities in these regions for the northern (southern) ring to be $n_{\mathrm{rim}} = 0.06_{-0.01}^{+0.01}~(0.07_{-0.01}^{+0.01})$ cm$^{-3}$ and $n_{\mathrm{out}} = 0.04_{-0.01}^{+0.01}~(0.04_{-0.01}^{+0.01})$ cm$^{-3}$. The best-fit models of the surface brightness profiles can be found in Figure \ref{fig:sbp}.

The temperature and density inside the brighter region are higher than those of the fainter surroundings. This confirms the nature of the discrepancy at $r_{\mathrm{rim}}\sim 1.3$ kpc as a shock driven by a nuclear outflow as mentioned by \citet{Machacek_2006b}. Utilising the $T_{\mathrm{rim}}/T_{\mathrm{out}}$ and $n_{\mathrm{rim}}/n_{\mathrm{out}}$ ratios in the Rankine-Hugoniot \citep[e.g.][]{Landau1959} shock conditions for a monatomic ideal gas with adiabatic index $\gamma = 5/3$ 
allows to estimate the speed of the shock's propagation.

Using the measured temperature discrepancies across each ring, we find the propagation speed of the shock for the northern (southern) ring to be $M_1 = 1.67 \pm 0.09$ ($M_1 = 1.74 \pm 0.09$), while the inferred density discrepancy measured through the change in surface brightness suggests a speed of $M_1 = 1.31 \pm 0.24$ ($M_1 = 1.28 \pm 0.23$). As stated in \citet{Machacek_2006b}, due to the narrowness of the shock front, the measured surface brightness discontinuity will underestimate the real density discrepancy. Therefore, this method infers the lower limits of the shock speed.

\subsubsection{The total energy of the outburst}
\label{section:Shock_propagation}

The speed of the shock front allows us to calculate its age and the power of the outburst that created it. For a mean radius of $R\sim0.81$ kpc estimated from the surface brightness profile, and the lower limit on the speed estimated from the density discontinuity, the age of the shock is $\sim1.8$ Myr, while the temperature change indicates a somewhat younger age of $\sim1.4$ Myr. The outburst energy can be expressed as $E_{\mathrm{shock}}\sim p_{\mathrm{rim}}V(p_{\mathrm{rim}}/p_{\mathrm{out}}-1)\sim 7.7\times 10^{54}$ erg, where $p_{\mathrm{rim}}$ and $p_{\mathrm{out}}$ represent the pressure of the shocked and unshocked gas. 

By combining the X-ray cavity volumes determined by \cite{Plsek2023} with the thermal pressure profiles from \cite{Plsek2022}, we estimated the total radio-mechanical energy released by the central AGN to $E_{\mathrm{cav}}=1.37_{-0.03}^{+0.02} \times 10^{55}$ erg. 
The total mechanical energy of the nuclear outburst is $E_{\mathrm{tot}} = E_{\mathrm{shock}} + E_{\mathrm{cav}}\sim 2.14\times 10^{55}$ erg. 

\section{Discussion}
\label{sec:sec5_discussion}


Super-Solar $\alpha$/Fe abundance ratios in the hot atmosphere of M89 would suggest that this galaxy is a rare case where the abundance ratios of the hot gas deviate from the Solar value, and are closer to the average stellar abundance ratios. Considering that this is potentially the first atmosphere with reported super-Solar abundance ratios and that M89 is a particularly dynamic system, our result requires careful investigation.

The stellar age of M89 is calculated to be $8.9_{-2.5}^{+3.4}$ Gyr \citep{McDermid_2006}, which corresponds to a formation time of $z \sim 0.7 - 4.3$. The early enrichment scenario, which is briefly explained in Section~\ref{sec:introduction}, suggests that the ICM gas has mostly completed its enrichment around $z \sim 2-3$. Assuming that the galaxy accumulated most of its hot X-ray emitting gas after the enrichment (i.e. after $z \sim 2-3$), M89 should have a hot atmosphere composed of `universal' gas. So far, other similar ellipticals have been found to host hot gas with Solar abundance ratios \citep[e.g.][]{Mernier2022b, Mernier2016a}, therefore it is critical to constrain the possible scenarios.

Observational indications for the super-Solar abundance ratios might be explained by the depletion of the original atmosphere due to AGN activity and ram-pressure stripping. M89 may have lost a significant part of the hot gas through turbulent mechanisms, replacing it with accumulated thermalised stellar wind wind material with super-Solar $\alpha$/Fe abundance ratios.


\subsection{Ram-pressure stripping and accretion cut-off}
\label{sec:Discuss1}



The tailored simulation studies of M89 by \cite{Roediger2015a, Roediger2015b} show that the observable X-ray tail itself is not the subject of stripping. Instead, it is a morphologically deformed atmosphere, shielded from the ICM and preserved by the galaxy up to or beyond pericenter passage. In more general terms, if the ambient medium of an infalling galaxy has sufficient viscosity and/or magnetic field, the Kelvin-Helmholtz instabilities that cause turbulent mixing are suppressed \citep{Chandrasekhar_1961}. Therefore, in an ICM that is viscous and has a strong enough magnetic field, the downstream hot atmosphere (i.e. the `tail') is unmixed, dense and therefore X-ray bright, which is believed to be the case of M89. This is confirmed by our measurements, as the elemental ratios of the Tail and Core regions are compatible, and the overall gas content in the tail is similar to that in the core. The less-dense Solar gas, which was once in the outskirts of the galaxy's original atmosphere, has been relocated at the edges of the tail due to the flow. As the simulations of M89 suggest, this less dense Solar gas at the edges of the tail is the subject of peeling. Consequently, as the motion continues, the Solar gas at the periphery constantly depletes. Moreover, the gas from the surrounding environment is no longer able to fall onto the galaxy because of the large velocity differences between the ICM and the galaxy \citep{Gunn1972}.




Before the infall, M89 experienced gas inflows and outflows simultaneously. At this unperturbed stage, the galaxy continuously accreted gas with Solar abundance ratios from the surrounding intergalactic medium, while part of its initial atmosphere was lost due to supernova-driven and AGN-induced outflows (for a review on the topic, see e.g. \citealp{Werner2018}). However, once the galaxy encounters the Virgo ICM, the accretion of external material stops, and the atmosphere begins to be stripped. During its oscillation inside the Virgo cluster, the thermalised stellar mass loss products, which have super-Solar abundance ratios are accumulated in the galaxy's atmosphere. As a result, the stellar mass loss products may become dominant in the hot atmosphere of M89.

We caution the reader that our proposed scenario is not compatible with a similar infalling galaxy NGC\,1404 which is also exhibiting an X-ray tail. \cite{Mernier2022b} showed that NGC\,1404, a satellite galaxy in the Fornax cluster, has Solar chemical composition. It is possible that the difference between the two stripped galaxies is a result of the AGN feedback in M89. Unlike M89 with AGN-induced X-ray cavities, NGC\,1404 does not exhibit any AGN activity. Therefore, the initial atmosphere of NGC\,1404 might not be uplifted and stripped as effectively. 

\subsection{Gas displacement by the AGN}

The outflow of the initial hot gas with Solar abundance ratios in M89 might be facilitated because of the AGN activity. In Section \ref{section:Shock_propagation}, we showed that the mechanical energy to evacuate both cavities is greater than 1.5$\times 10^{55}$ erg. The total mechanical energy of the nuclear outburst is $\sim 2.14\times 10^{55}$ erg, and the mechanical power of the nuclear outburst is $1.3 - 1.7 \times 10^{41}$ erg s$^{-1}$. For comparison, the gravitational binding energy of the galaxy is calculated as $E_{\rm bind} = \sim 3.97 \times 10^{56}~{\rm erg}$ within a half-light radius ($r_{\rm e}$), using the total mass from \cite{Cappellari2013} and X-ray gas mass from \cite{Plsek2022}.

Based on these results, the energy released into the galaxy atmosphere from the AGN-induced outflows in M89 is significant. As the atmospheric gas is uplifted to higher altitudes, ram-pressure stripping becomes more effective, facilitating the loss of the original galactic atmosphere.

\subsection{Replenishment due to the stellar mass loss}
\label{sec:Discuss2}

Internal sources of hot gas in elliptical galaxies are the thermalised stellar mass loss products \citep{Mathews1990, MathewsBrighetni2003}. When the red giant stars orbit supersonically relative to the hot ambient gas, simulations suggest that approximately 75\% of the ejected gas is shock-heated and becomes part of the hot atmosphere \citep{Parriott_2008}. Planetary nebulae are another source of hot gas, although the interaction between the hot ambient medium and a planetary nebula shells thermalizes only a small fraction of the material \citep{Bregman2009}. 

In the core of M89, due to the lack of ongoing Solar gas accretion, the internal sources of hot gas can dominate the overall chemical composition. The presence of higher-than-average SNcc products in the stars of the galaxy \citep{Lonoce2021} supports this scenario. In this section, we focus on the ejected wind material of evolving stars and compare it to the metal masses in the hot gas based on our measured metal abundances.


\subsubsection{Total mass budget of metals}

We find the individual metal budget in the hot gas phase. The total X-ray emitting hot gas in M89 is derived as $\sim1.8 \times 10^{8}\,M_{\odot}$ for 2 $r_{\text{e}}$ region \citep{Su_2015}. The elemental metal mass in a hot gas region enclosed with radius $R$ can be calculated as
\begin{equation}
    M_{X}(R) = X_{\odot} X(R) \frac{A_{X}}{A_{H}} \rchi_{H} M_{\text{gas}}(R),
        \label{eq:metal_mass}
\end{equation}
where $X_{\odot}$ and $X(R)$ are the Solar abundance of element X, taken from \citet{Lodders2009}, and the measured abundance of that element, respectively. Their product, $(N_{X\odot}/N_{H\odot}) \times (N_{X}/N_{H}) / (N_{X\odot}/N_{H\odot})$, gives the ratio of number of $X$ atoms over $H$ atoms in the medium. $A_{X}$ and $A_{H}$ are the atomic weights of X and H respectively. The obtained mass fraction of element $X$ with respect to $H$ is then multiplied by the $\rchi_{H}$, the $H$ mass fraction adopted as 0.735, which gives the ratio of X mass over the total mass. $M_{\text{gas}}(R)$ is the total mass in the enclosed volume, consequently, their product gives the mass of element $X$. Finally, the calculated hot-gas phase metals in the centre of M89 are presented in Table \ref{table:masses}.

\setlength{\tabcolsep}{7.8pt}
\renewcommand{\arraystretch}{1.65}
    \begin{table}
\caption{Masses of the gas-phase metals inside M89 at $2\,r_{\text{e}}$ obtained from the abundance values from the Core region. }
\centering
\begin{tabular}{lccc|ccc|ccc}
   \hline
   Element & EPIC ($10^5 M_{\odot}$) & RGS ($10^5 M_{\odot}$)  & ACIS-S ($10^5 M_{\odot}$) \\
    \hline
    O  & --                   & $3.71\pm1.02$ & --                   \\
    Ne & --                   & $1.10^{+0.40}_{-0.36}$ & $6.16^{+1.22}_{-1.06}$ \\
    Mg & $0.85\pm0.18$ & $0.64\pm0.20$ & $2.01^{+0.52}_{-0.32}$ \\
    Si & $0.78^{+0.20}_{-0.17}$ & --                   & $1.69^{+0.44}_{-0.31}$ \\
    S  & $0.93\pm0.31$          & --                   & $1.07^{+0.48}_{-0.33}$ \\
    Fe & $1.50^{+0.19}_{-0.17}$        & $0.55\pm0.09$ & $2.51^{+0.72}_{-0.65}$\\
    \hline
\end{tabular}
\label{table:masses}
\end{table}

\begin{figure*}
    \centering
        \includegraphics[width=0.32\textwidth]{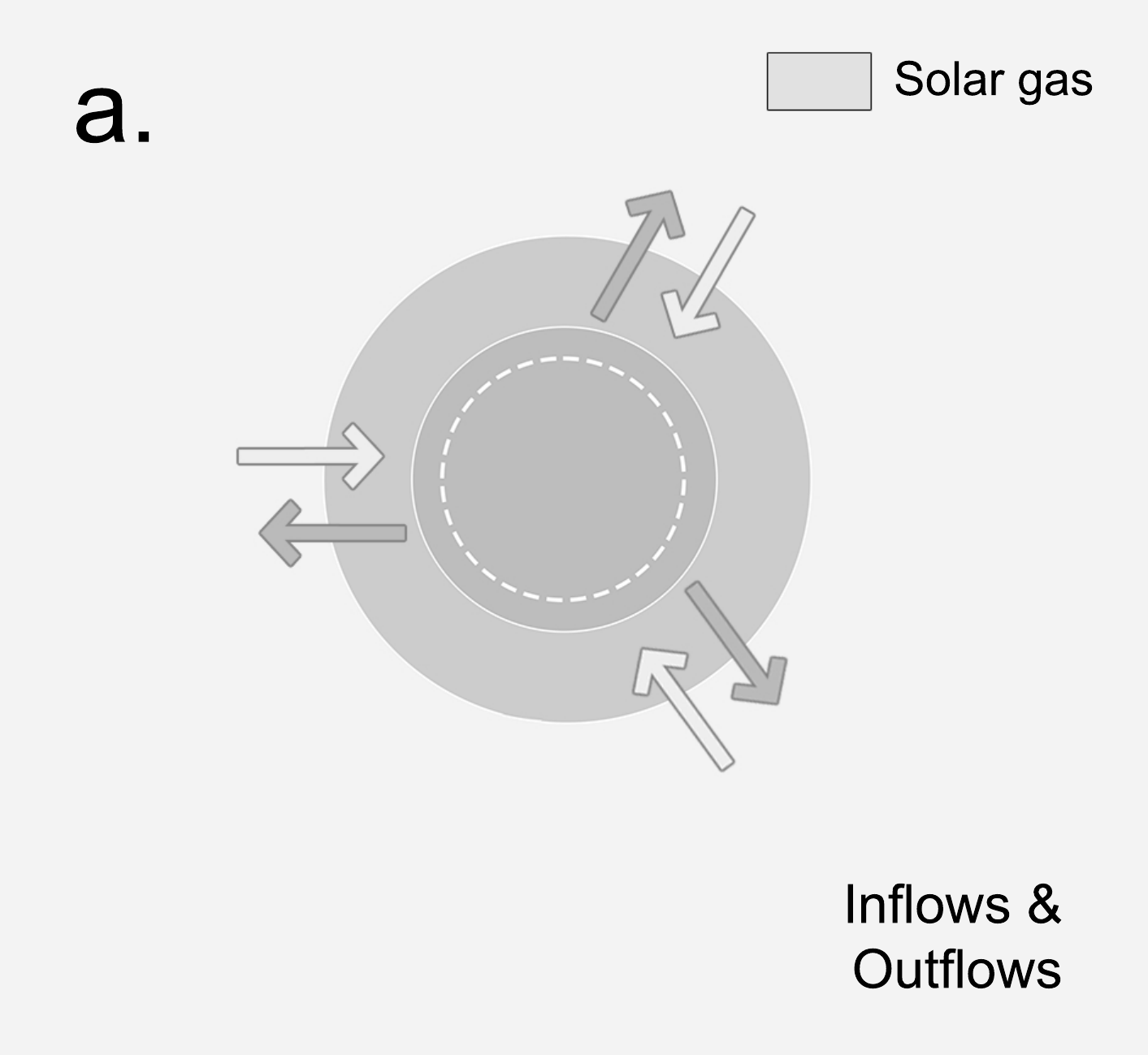} \hfill
    \includegraphics[width=0.32\textwidth]{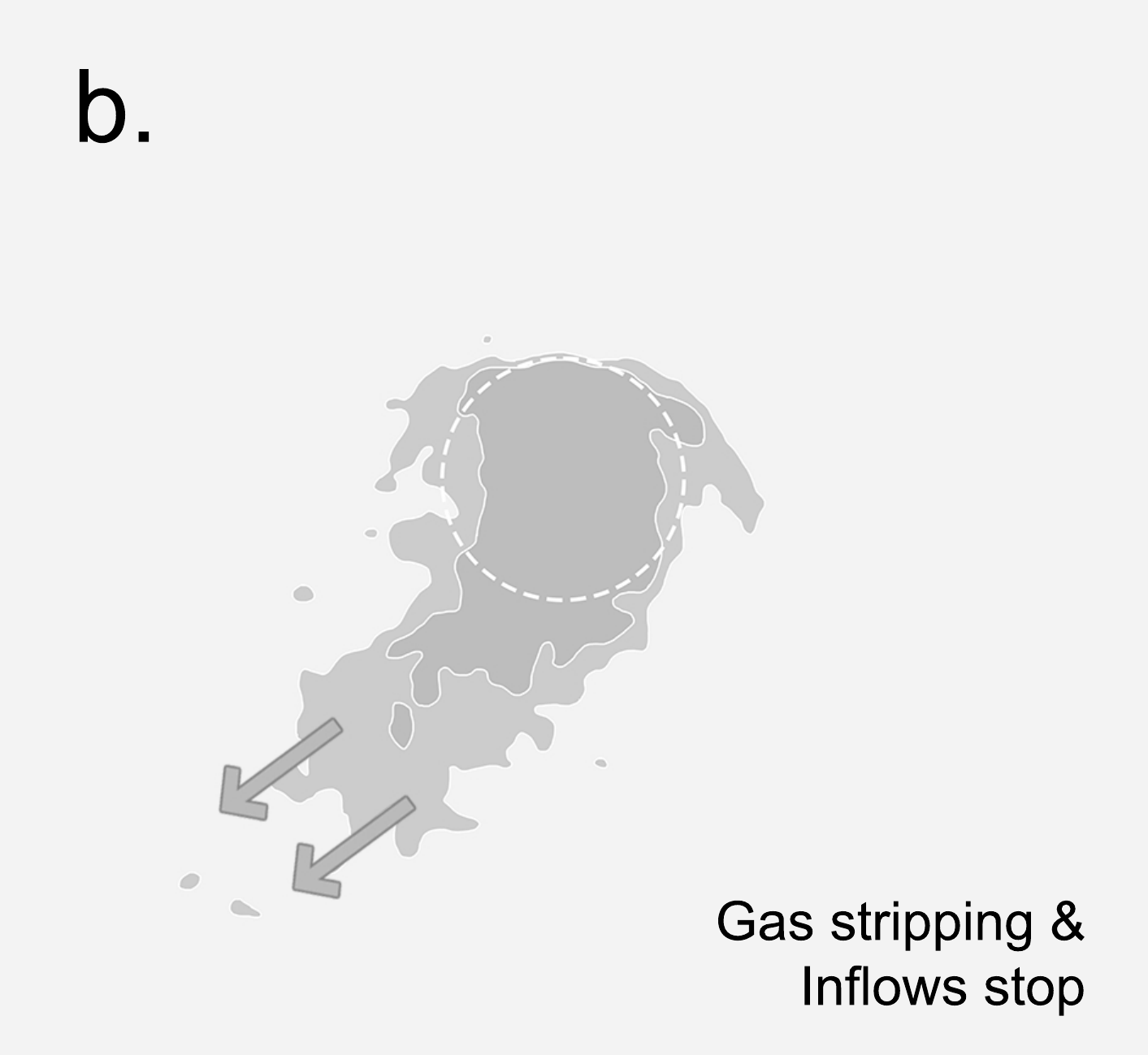}
    \hfill
    \includegraphics[width=0.32\textwidth]{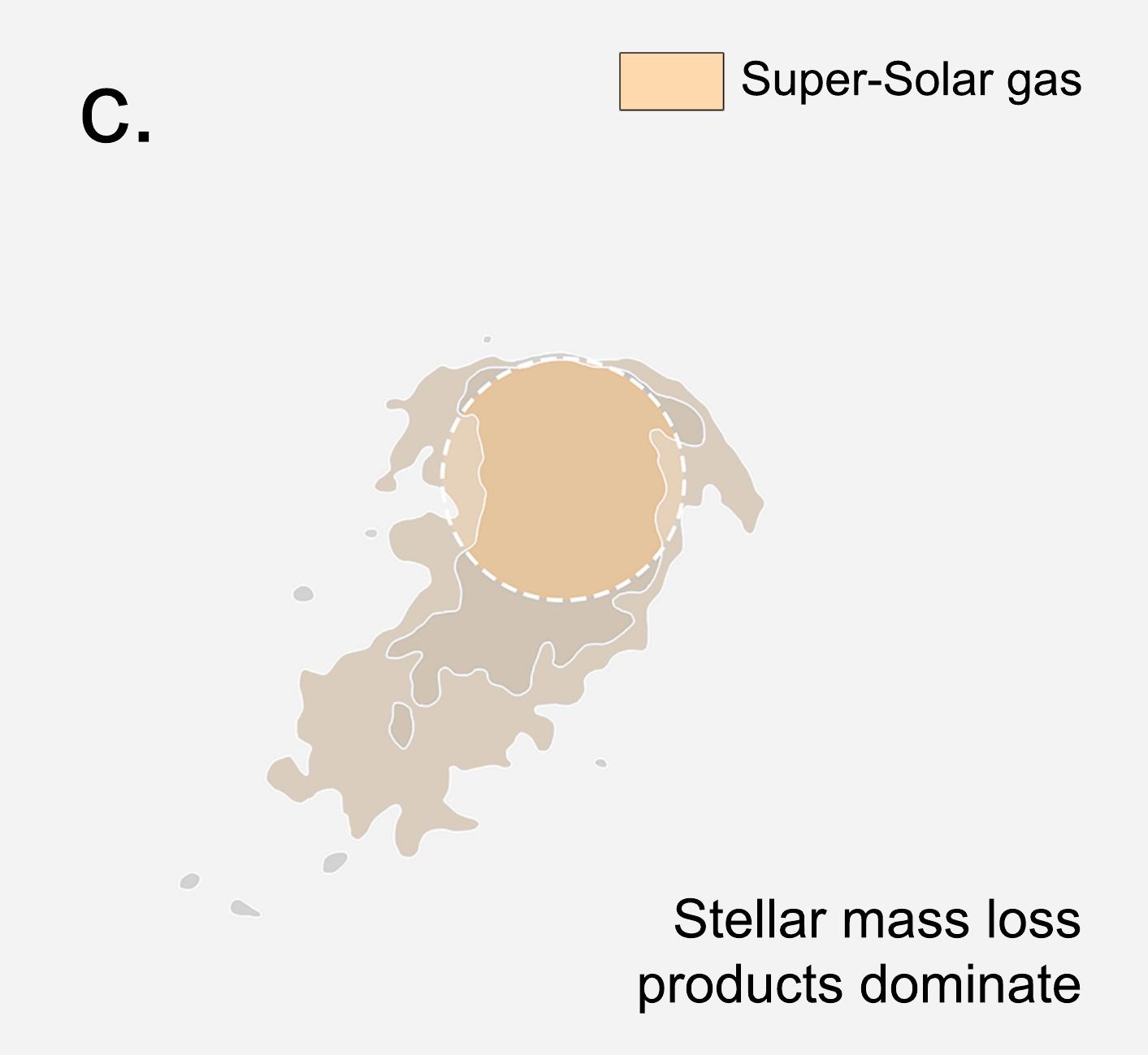} \\
    \caption{Illustration of the proposed chemical enrichment scenario of M89. The colour shade changes in the illustrative sketch are arbitrary and present density changes in a simplified approach, where the lighter outskirts have high entropy and darker inner regions have low entropy. The initial size of the hot atmosphere is again arbitrary, while the dashed circle corresponds to the Core region.  (\textbf{a.}) Continuous inflows and outflows occur between the M89 and the intergalactic medium. The X-ray gas in both mediums has Solar abundance ratios. (\textbf{b.}) Once the galaxy falls into the Virgo cluster, because of the ram pressure and AGN activity, its original atmosphere begins to be stripped.  Due to the accretion cut-off, the galaxy is not able to accrete new gas from the surrounding ICM. (\textbf{c.}) As the inflow of gas is prevented and the stellar content of the galaxy ejects fresh gas with super-Solar abundance ratios via the stellar winds, the super-Solar gas becomes dominant.}
    \label{fig:sketch}
\end{figure*}

\subsubsection{Stellar wind contribution to the hot gas}

The stellar mass loss rate  in an elliptical galaxy with a Salpeter initial mass function \citep{Salpater1955} in a given time $t$, can be approximated according to \citep{Ciotti} as following:
\begin{equation}
    \dot{M}_{*}(t) = 1.5 \times 10^{-11}L_{\text{B}}  (t / t_{H})^{-1.3}
\end{equation}
where $L_{B}$ is the present B band luminosity and $t_{H}$ is the Hubble time. From \citet{Ellis2006}, we adopt the $L_{B} = 1.95 \times 10^{10} L_{\odot}$ for M89. As for determining the time interval for the integral, we investigate the star formation of the galaxy. Using MegaCam observations,  \citet{Boselli_2022} showed that the $\mathrm{H\alpha[NII]}$ distribution in M89 mostly comes from the diffuse filamentary structures rather than any apparent star-forming region. Moreover, HST observations of the galaxy do not show significant $\mathrm{H\alpha[NII]}$ emission \citep{Temi_2022}. Therefore, similar to other elliptical galaxies, we assume that the galaxy experienced a short and brief star formation peak around $z \approx 3$ \citep{Thomas2010}. As a result, we integrate the mass-loss equation from $z = 3$  to $z = 0$ ($t = t_{H}$), and find a total mass of $10.6 \times 10^{9} \, M_{\odot}$. Notice that this approximation is very rough and should be taken as an order-of-magnitude estimation for further analysis.

For the next step, we take the average ranges of stellar abundances in M89 from \citet{Lonoce2021} and calculate the individual metal masses. Using Equation \ref{eq:metal_mass}, the ejected masses via stellar wind are found to be $(1.2 - 1.6) \times 10^{7} \, M_{\odot}$ for Fe, $(1.1 - 1.7) \times 10^{7} \, M_{\odot}$ for Mg and $(1.6 - 2.1) \times 10^{7} \, M_{\odot}$ for O.

Note that the above calculations are rough predictions for order of magnitude comparison between the observed and theoretically ejected metal masses. Nonetheless, we find 1--2 orders of magnitude more mass is ejected through stellar winds compared to the total observed mass of O, Mg and Fe. Therefore it is possible to say that the hot gas with super-Solar ratios could originate from the super-Solar stellar content present in M89.

In this scenario, gas outflows and inflows existed simultaneously before the infall (see (a.) in Figure \ref{fig:sketch}). After the infall, the original Solar gas content of M89 begins to be depleted because of the ram-pressure stripping and AGN-induced outflows. Furthermore, during the infall, accretion of the Solar gas from surroundings is restrained (see (b.) in Figure \ref{fig:sketch}). The stars in M89, which have a higher $\alpha$/Fe ratio than the average elliptical galaxies, eject hot gas into the galaxy via stellar winds. As a result, we argue that the freshly ejected gas is mixed with the remaining original atmosphere and creates a new generation of hot gas with super-Solar abundance ratios (see (c.) in Figure \ref{fig:sketch}).

We note that \citet{Mernier2022b} showed that the metal mass produced and ejected through stars in NGC 1404 also exceeds the observed mass: $\sim$20 and $\sim$40 times more mass is produced for Fe and Mg, respectively. Interestingly, the hot atmosphere of NGC 1404 is found to be Solar by that study. Therefore, any metal excess caused by stellar winds does not necessarily indicate a super-Solar composition in an X-ray gas. By studying the differences between two ram-pressure stripped galaxies NGC\,1404 (which does not harbour an active nucleus) and M89 (which harbours a prominent AGN with bubbles and shocks), we can attempt to derive constraints on the chemical enrichment and the origin of galactic atmospheres. We can address questions such as: (i) is a stellar-like chemical composition in a hot gas a universal feature in galaxies with an accretion cut-off? (ii) Why does NGC\,1404 have Solar ratios while M89 has super-Solar ratios? More generally, (iii) is it enough to experience an accretion cut-off to have a stellar-like atmosphere, or is it necessary to undergo a significant loss of original gas? We conclude that, in M89, AGN-induced outflows could have facilitated the stripping of the original galactic atmosphere, which has been replaced with fresh stellar mass loss material with super-Solar $\alpha$/Fe abundance ratios, producing the observed difference between M89 and NGC\,1404.

Moreover, the metal budget in the ICM gas is measured to be 2 to 10 times larger than what their stars could have produced (e.g. \citealp{Renzini_2014}). Similar to the \citet{Mernier2022b}, we observe an `inverse metal conundrum' in M89 where the metals in the hot gas are $1-2$ orders of magnitude less than the stars could have ejected. The two conundra might be comparable. It is possible that the excess metals that are ejected by the stars of elliptical galaxies are mixed with the ICM during the enrichment.

\section{Conclusions}
\label{sec:sec6_conclusion}

In this paper, we investigated the chemical composition of the infalling elliptical galaxy M89 (NGC\,4552), which hosts X-ray cavities and a prominent shock front due to radio mechanical AGN activity. During its voyage into the Virgo cluster, the hot galaxy atmosphere experiences ram-pressure stripping caused by the interaction with the intracluster medium. Due to the stripping, M89 hosts a bright X-ray tail. 


Using \textit{XMM-Newton} and \textit{Chandra} archival observations, we investigated the chemical composition in the core and the tail of the galaxy. With particular attention to the modelling, we derived O/Fe, Ne/Fe, Mg/Fe, Si/Fe and S/Fe ratios from EPIC MOS, EPIC pn, RGS and ACIS-S. Our results can be summarised as follows.

\begin{itemize}
  \item Our measurements suggest a super-Solar abundance ratios (i.e. $\alpha$/Fe > 1) in the core. Such a gas content is chemically closer to the stellar population of the galaxy \citep{Lonoce2021} rather than the rest of the gaseous content in the universe, which is closer to our Solar system (i.e. $\alpha$/Fe $\sim$ 1). In the tail, the results are comparable with the core, however, it is less certain due to large uncertainties.
  
 \item We report a fitting bias in the \textit{XMM-Newton}/RGS data. In low-temperature plasma, the O/Fe ratio changes significantly (> 1$\sigma$) between different multi-temperature models. Because of the bias, we were unable to robustly constrain the SNcc contribution fraction as there is a systematic change in the SNcc ratios. Nevertheless, we found that the SNcc contribution to the hot atmosphere of M89 is greater than 70\% for all multi-temperature models and instruments.

  \item We argue that the potential super-Solar abundance ratios might be produced mainly by the stellar population of the galaxy via stellar winds after the infall. During the infall, the low-density Solar gas is stripped from the edges of the tail. Moreover, because of the motion of the galaxy, the continual infall of gas with Solar abundance ratios from the surrounding ICM is stopped. Therefore, a new generation of hot atmosphere with super-Solar abundance ratios might be produced due to the stellar mass loss products in M89. 
  
  \item In order to test the stellar contribution to the observed chemical composition, we calculated the mass of the observed hot gas metals and compared them to the metal content that stellar winds possibly eject since $z = 3$. We showed that the stellar winds alone could possibly produce $\sim2$ orders of magnitude more metal mass than the metal budget of the hot gas in M89.

  \item In the AGN activity analysis, we showed that the AGN-induced nuclear outburst energy is $\sim2.14\times 10^{55}$ erg; while the gravitational binding energy is $\sim3.97 \times 10^{56}~{\rm erg}$. We conclude that the comparable energies suggest that AGN significantly facilitates the stripping of the original galaxy atmosphere.
  
  \item We state that the chemical enrichment history of galaxies is still a matter of debate. Based on our results, we argue a possible scenario that would explain the potential super-Solar abundance ratios. However, this scenario does not explain such a composition universally. For instance, an atmosphere of similar infalling galaxy NGC\,1404 exhibits Solar abundance ratios. We show that in M89 the AGN activity might facilitate ram-pressure stripping, possibly clarifying the differences between the two galaxies.  
\end{itemize}

This work shows the importance of studying low-mass systems in various dynamical states to improve our understanding of the chemical enrichment of the universe. Due to the relatively short exposure time of \textit{XMM-Newton}, we were unable to investigate the spatial distribution of metals. In order to fully understand the metal mixing between the X-ray tail and the Virgo ICM, as well as the AGN-induced metal distribution in the galaxy core, a deeper observation by \textit{XMM-Newton} is required. Future missions with higher spectral resolution, such as \textit{XRISM} and \textit{Athena}, will help us to significantly improve the systematic uncertainties in the measurements of the abundances of SN$_{\rm CC}$ products such as O, Ne, and Mg.

\section*{Acknowledgements}

The authors thank the anonymous referee for constructive feedback that helped improve this paper. S.K. and E.N.E. would like to thank TUBITAK for the financial support under the 1002 project with code number 121F436. The research leading to these results has received funding from the European Union’s Horizon 2020 Programme under the AHEAD2020 project (grant agreement n. 871158). T.P., N.W. and J-P.B. were supported by GACR grant 21-13491X. The material is based upon work supported by NASA under award number 80GSFC21M0002. E.N.E. would like to thank Bogazici University for their support
through BAP project number 13760. This work is based on observations obtained with \textit{XMM-Newton}, an ESA science mission with instruments and contributions directly funded by ESA member states and the USA (NASA). The scientific results reported in this article are based in part on data obtained from the \textit{Chandra} Data Archive.

\section*{Data Availability}

The original data discussed in this article can be accessed publicly from the \textit{XMM-Newton} Science Archive (https://nxsa.esac.esa.int/nxsa-web/) and the \textit{Chandra} X-ray Center (https://cda.harvard.edu/chaser/). The \textit{XMM-Newton} and \textit{Chandra} data are processed using the \code{SAS} software (https://www.cosmos.esa.int/web/xmm-newton/sas) and \code{CIAO} software (https://cxc.cfa.harvard.edu/ciao), respectively. Additional derived data products can be obtained from the main author upon request.



\bibliographystyle{mnras}
\bibliography{example} 

\begin{thebibliography}{}
\makeatletter
\relax
\def\mn@urlcharsother{\let\do\@makeother \do\$\do\&\do\#\do\^\do\_\do\%\do\~}
\def\mn@doi{\begingroup\mn@urlcharsother \@ifnextchar [ {\mn@doi@} {\mn@doi@[]}}
\def\mn@doi@[#1]#2{\def\@tempa{#1}\ifx\@tempa\@empty \href {http://dx.doi.org/#2} {doi:#2}\else \href {http://dx.doi.org/#2} {#1}\fi \endgroup}
\def\mn@eprint#1#2{\mn@eprint@#1:#2::\@nil}
\def\mn@eprint@arXiv#1{\href {http://arxiv.org/abs/#1} {{\tt arXiv:#1}}}
\def\mn@eprint@dblp#1{\href {http://dblp.uni-trier.de/rec/bibtex/#1.xml} {dblp:#1}}
\def\mn@eprint@#1:#2:#3:#4\@nil{\def\@tempa {#1}\def\@tempb {#2}\def\@tempc {#3}\ifx \@tempc \@empty \let \@tempc \@tempb \let \@tempb \@tempa \fi \ifx \@tempb \@empty \def\@tempb {arXiv}\fi \@ifundefined {mn@eprint@\@tempb}{\@tempb:\@tempc}{\expandafter \expandafter \csname mn@eprint@\@tempb\endcsname \expandafter{\@tempc}}}

\bibitem[\protect\citeauthoryear{{Arnaud}}{{Arnaud}}{1996}]{Arnaud1996}
{Arnaud} K.~A.,  1996, in {Jacoby} G.~H.,  {Barnes} J.,  eds,  Astronomical Society of the Pacific Conference Series Vol. 101, Astronomical Data Analysis Software and Systems V. p.~17

\bibitem[\protect\citeauthoryear{{Berkhuijsen}}{{Berkhuijsen}}{1971}]{Berkhuijsen_1971}
{Berkhuijsen} E.~M.,  1971, \aap, \href {https://ui.adsabs.harvard.edu/abs/1971A&A....14..359B} {14, 359}

\bibitem[\protect\citeauthoryear{Biffi et~al.,}{Biffi et~al.}{2017}]{Biffi2017}
Biffi V.,  et~al., 2017, \mn@doi [\mnras] {10.1093/mnras/stx444}, 468, 531

\bibitem[\protect\citeauthoryear{{Boselli, A.} et~al.,}{{Boselli, A.} et~al.}{2022}]{Boselli_2022}
{Boselli, A.} et~al., 2022, \mn@doi [A\&A] {10.1051/0004-6361/202142482}, 659, A46

\bibitem[\protect\citeauthoryear{{Bregman} \& {Parriott}}{{Bregman} \& {Parriott}}{2009}]{Bregman2009}
{Bregman} J.~N.,  {Parriott} J.~R.,  2009, \mn@doi [\apj] {10.1088/0004-637X/699/2/923}, \href {https://ui.adsabs.harvard.edu/abs/2009ApJ...699..923B} {699, 923}

\bibitem[\protect\citeauthoryear{{Buote}}{{Buote}}{2000}]{Buote2000}
{Buote} D.~A.,  2000, \mn@doi [\apj] {10.1086/309224}, \href {https://ui.adsabs.harvard.edu/abs/2000ApJ...539..172B} {539, 172}

\bibitem[\protect\citeauthoryear{{Buote} \& {Fabian}}{{Buote} \& {Fabian}}{1998}]{Buote1998}
{Buote} D.~A.,  {Fabian} A.~C.,  1998, \mn@doi [\mnras] {10.1046/j.1365-8711.1998.01478.x}, \href {https://ui.adsabs.harvard.edu/abs/1998MNRAS.296..977B} {296, 977}

\bibitem[\protect\citeauthoryear{{Cappellari} et~al.,}{{Cappellari} et~al.}{2013}]{Cappellari2013}
{Cappellari} M.,  et~al., 2013, \mn@doi [\mnras] {10.1093/mnras/stt562}, \href {https://ui.adsabs.harvard.edu/abs/2013MNRAS.432.1709C} {432, 1709}

\bibitem[\protect\citeauthoryear{{Cash}}{{Cash}}{1979}]{Cash1979}
{Cash} W.,  1979, \mn@doi [\apj] {10.1086/156922}, \href {https://ui.adsabs.harvard.edu/abs/1979ApJ...228..939C} {228, 939}

\bibitem[\protect\citeauthoryear{{Chandrasekhar}}{{Chandrasekhar}}{1961}]{Chandrasekhar_1961}
{Chandrasekhar} S.,  1961, {Hydrodynamic and hydromagnetic stability}.
{Oxford University Press}

\bibitem[\protect\citeauthoryear{{Ciotti}, {D'Ercole}, {Pellegrini}  \& {Renzini}}{{Ciotti} et~al.}{1991}]{Ciotti}
{Ciotti} L.,  {D'Ercole} A.,  {Pellegrini} S.,   {Renzini} A.,  1991, \mn@doi [\apj] {10.1086/170289}, \href {https://ui.adsabs.harvard.edu/abs/1991ApJ...376..380C} {376, 380}

\bibitem[\protect\citeauthoryear{Conroy, Graves  \& van Dokkum}{Conroy et~al.}{2013}]{Conroy_2013}
Conroy C.,  Graves G.~J.,   van Dokkum P.~G.,  2013, \mn@doi [\apj] {10.1088/0004-637x/780/1/33}, 780, 33

\bibitem[\protect\citeauthoryear{{De Luca} \& {Molendi}}{{De Luca} \& {Molendi}}{2004}]{deLuca_2004}
{De Luca} {Molendi} 2004, \mn@doi [A\&A] {10.1051/0004-6361:20034421}, 419, 837

\bibitem[\protect\citeauthoryear{{Ellis} \& {O'Sullivan}}{{Ellis} \& {O'Sullivan}}{2006}]{Ellis2006}
{Ellis} S.~C.,  {O'Sullivan} E.,  2006, \mn@doi [\mnras] {10.1111/j.1365-2966.2005.09982.x}, \href {https://ui.adsabs.harvard.edu/abs/2006MNRAS.367..627E} {367, 627}

\bibitem[\protect\citeauthoryear{Erdim, Ezer, Ünver, Hazar  \& Hudaverdi}{Erdim et~al.}{2021}]{Erdim_2021}
Erdim M.~K.,  Ezer C.,  Ünver O.,  Hazar F.,   Hudaverdi M.,  2021, \mn@doi [\mnras] {10.1093/mnras/stab2730}, 508, 3337

\bibitem[\protect\citeauthoryear{{Filho}, {Fraternali}, {Markoff}, {Nagar}, {Barthel}, {Ho}  \& {Yuan}}{{Filho} et~al.}{2004}]{Filho2004}
{Filho} M.~E.,  {Fraternali} F.,  {Markoff} S.,  {Nagar} N.~M.,  {Barthel} P.~D.,  {Ho} L.~C.,   {Yuan} F.,  2004, \mn@doi [\aap] {10.1051/0004-6361:20034486}, \href {https://ui.adsabs.harvard.edu/abs/2004A&A...418..429F} {418, 429}

\bibitem[\protect\citeauthoryear{Fink et~al.,}{Fink et~al.}{2014}]{Fink_2014}
Fink M.,  et~al., 2014, \mn@doi [\mnras] {10.1093/mnras/stt2315}, 438, 1762

\bibitem[\protect\citeauthoryear{{Foster}, {Ji}, {Smith}  \& {Brickhouse}}{{Foster} et~al.}{2012}]{Foster2012}
{Foster} A.~R.,  {Ji} L.,  {Smith} R.~K.,   {Brickhouse} N.~S.,  2012, \mn@doi [\apj] {10.1088/0004-637X/756/2/128}, \href {https://ui.adsabs.harvard.edu/abs/2012ApJ...756..128F} {756, 128}

\bibitem[\protect\citeauthoryear{{Fruscione} et~al.,}{{Fruscione} et~al.}{2006}]{Fruscione2006}
{Fruscione} A.,  et~al., 2006, in {Silva} D.~R.,  {Doxsey} R.~E.,  eds,  Society of Photo-Optical Instrumentation Engineers (SPIE) Conference Series Vol. 6270, Society of Photo-Optical Instrumentation Engineers (SPIE) Conference Series. p. 62701V, \mn@doi{10.1117/12.671760}

\bibitem[\protect\citeauthoryear{{Fukushima}, {Kobayashi}  \& {Matsushita}}{{Fukushima} et~al.}{2022}]{Fukushima2022}
{Fukushima} K.,  {Kobayashi} S.~B.,   {Matsushita} K.,  2022, \mn@doi [\mnras] {10.1093/mnras/stac1590}, \href {https://ui.adsabs.harvard.edu/abs/2022MNRAS.514.4222F} {514, 4222}

\bibitem[\protect\citeauthoryear{{Gastaldello}, {Simionescu}, {Mernier}, {Biffi}, {Gaspari}, {Sato}  \& {Matsushita}}{{Gastaldello} et~al.}{2021}]{Gastaldello2021}
{Gastaldello} F.,  {Simionescu} A.,  {Mernier} F.,  {Biffi} V.,  {Gaspari} M.,  {Sato} K.,   {Matsushita} K.,  2021, \mn@doi [Universe] {10.3390/universe7070208}, \href {https://ui.adsabs.harvard.edu/abs/2021Univ....7..208G} {7, 208}

\bibitem[\protect\citeauthoryear{Gatuzz et~al.,}{Gatuzz et~al.}{2023}]{Gatuzz2023}
Gatuzz E.,  et~al., 2023, \mn@doi [\mnras] {10.1093/mnras/stad447}, 520, 4793

\bibitem[\protect\citeauthoryear{{Gunn} \& {Gott}}{{Gunn} \& {Gott}}{1972a}]{Gunn_1972}
{Gunn} J.~E.,  {Gott} J.~Richard I.,  1972a, \mn@doi [\apj] {10.1086/151605}, 176, 1

\bibitem[\protect\citeauthoryear{{Gunn} \& {Gott}}{{Gunn} \& {Gott}}{1972b}]{Gunn1972}
{Gunn} J.~E.,  {Gott} J.~Richard I.,  1972b, \mn@doi [\apj] {10.1086/151605}, \href {https://ui.adsabs.harvard.edu/abs/1972ApJ...176....1G} {176, 1}

\bibitem[\protect\citeauthoryear{{HI4PI Collaboration} et~al.,}{{HI4PI Collaboration} et~al.}{2016}]{HI4PI2016}
{HI4PI Collaboration} et~al., 2016, \mn@doi [\aap] {10.1051/0004-6361/201629178}, \href {https://ui.adsabs.harvard.edu/abs/2016A&A...594A.116H} {594, A116}

\bibitem[\protect\citeauthoryear{{Hickox} \& {Alexander}}{{Hickox} \& {Alexander}}{2018}]{Hickox2018}
{Hickox} R.~C.,  {Alexander} D.~M.,  2018, \mn@doi [\araa] {10.1146/annurev-astro-081817-051803}, \href {https://ui.adsabs.harvard.edu/abs/2018ARA&A..56..625H} {56, 625}

\bibitem[\protect\citeauthoryear{{Hitomi Collaboration}}{{Hitomi Collaboration}}{2017}]{Hitomi2017}
{Hitomi Collaboration} 2017, \mn@doi [Nature] {10.1038/nature24301}, 551, 478

\bibitem[\protect\citeauthoryear{Irwin, Athey  \& Bregman}{Irwin et~al.}{2003}]{Irwin_2003}
Irwin J.~A.,  Athey A.~E.,   Bregman J.~N.,  2003, \mn@doi [\apj] {10.1086/368179}, 587, 356

\bibitem[\protect\citeauthoryear{{Ji}, {Irwin}, {Athey}, {Bregman}  \& {Lloyd-Davies}}{{Ji} et~al.}{2009}]{Ji2009}
{Ji} J.,  {Irwin} J.~A.,  {Athey} A.,  {Bregman} J.~N.,   {Lloyd-Davies} E.~J.,  2009, \mn@doi [\apj] {10.1088/0004-637X/696/2/2252}, \href {https://ui.adsabs.harvard.edu/abs/2009ApJ...696.2252J} {696, 2252}

\bibitem[\protect\citeauthoryear{{Kaastra} \& {Bleeker}}{{Kaastra} \& {Bleeker}}{2016}]{Kaastra_2016}
{Kaastra} {Bleeker} 2016, \mn@doi [A\&A] {10.1051/0004-6361/201527395}, 587, A151

\bibitem[\protect\citeauthoryear{Landau \& Lifshitz}{Landau \& Lifshitz}{1959}]{Landau1959}
Landau L.,  Lifshitz E.,  1959, Fluid Mechanics.
Pergamon Press

\bibitem[\protect\citeauthoryear{{Lodders}, {Palme}  \& {Gail}}{{Lodders} et~al.}{2009}]{Lodders2009}
{Lodders} K.,  {Palme} H.,   {Gail} H.~P.,  2009, \mn@doi [Landolt Börnstein] {10.1007/978-3-540-88055-4_34}, \href {https://ui.adsabs.harvard.edu/abs/2009LanB...4B..712L} {4B, 712}

\bibitem[\protect\citeauthoryear{Lonoce, Feldmeier-Krause  \& Freedman}{Lonoce et~al.}{2021}]{Lonoce2021}
Lonoce I.,  Feldmeier-Krause A.,   Freedman W.~L.,  2021, \mn@doi [\apj] {10.3847/1538-4357/ac11f9}, 920, 93

\bibitem[\protect\citeauthoryear{Machacek, Jones, Forman  \& Nulsen}{Machacek et~al.}{2006a}]{Machacek_2006a}
Machacek M.,  Jones C.,  Forman W.~R.,   Nulsen P.,  2006a, \mn@doi [\apj] {10.1086/503350}, 644, 155–166

\bibitem[\protect\citeauthoryear{Machacek, Nulsen, Jones  \& Forman}{Machacek et~al.}{2006b}]{Machacek_2006b}
Machacek M.,  Nulsen P. E.~J.,  Jones C.,   Forman W.~R.,  2006b, \mn@doi [\apj] {10.1086/505963}, 648, 947–955

\bibitem[\protect\citeauthoryear{Madau \& Dickinson}{Madau \& Dickinson}{2014}]{MadauDickinson2014}
Madau P.,  Dickinson M.,  2014, \mn@doi [\araa] {10.1146/annurev-astro-081811-125615}, 52, 415

\bibitem[\protect\citeauthoryear{{Mathews}}{{Mathews}}{1990}]{Mathews1990}
{Mathews} W.~G.,  1990, \mn@doi [\apj] {10.1086/168708}, \href {https://ui.adsabs.harvard.edu/abs/1990ApJ...354..468M} {354, 468}

\bibitem[\protect\citeauthoryear{Mathews \& Brighenti}{Mathews \& Brighenti}{2003}]{MathewsBrighetni2003}
Mathews W.~G.,  Brighenti F.,  2003, \mn@doi [\araa] {10.1146/annurev.astro.41.090401.094542}, 41, 191

\bibitem[\protect\citeauthoryear{{Matsushita}, {B{\"o}hringer}, {Takahashi}  \& {Ikebe}}{{Matsushita} et~al.}{2007}]{Matsushita2007}
{Matsushita} K.,  {B{\"o}hringer} H.,  {Takahashi} I.,   {Ikebe} Y.,  2007, \mn@doi [\aap] {10.1051/0004-6361:20041577}, \href {https://ui.adsabs.harvard.edu/abs/2007A&A...462..953M} {462, 953}

\bibitem[\protect\citeauthoryear{McDermid et~al.,}{McDermid et~al.}{2006}]{McDermid_2006}
McDermid R.~M.,  et~al., 2006, \mn@doi [\mnras] {10.1111/j.1365-2966.2006.11065.x}, 373, 906

\bibitem[\protect\citeauthoryear{{Mernier} \& {Biffi}}{{Mernier} \& {Biffi}}{2022}]{Mernier2022}
{Mernier} F.,  {Biffi} V.,  2022, arXiv e-prints, \href {https://ui.adsabs.harvard.edu/abs/2022arXiv220207097M} {p. arXiv:2202.07097}

\bibitem[\protect\citeauthoryear{{Mernier}, {de Plaa}, {Lovisari}, {Pinto}, {Zhang}, {Kaastra}, {Werner}  \& {Simionescu}}{{Mernier} et~al.}{2015b}]{Mernier2015}
{Mernier} F.,  {de Plaa} J.,  {Lovisari} L.,  {Pinto} C.,  {Zhang} Y.~Y.,  {Kaastra} J.~S.,  {Werner} N.,   {Simionescu} A.,  2015b, \mn@doi [\aap] {10.1051/0004-6361/201425282}, \href {https://ui.adsabs.harvard.edu/abs/2015A&A...575A..37M} {575, A37}

\bibitem[\protect\citeauthoryear{{Mernier}, {de Plaa, J.}, {Lovisari, L.}, {Pinto, C.}, {Zhang, Y.-Y.}, {Kaastra, J. S.}, {Werner, N.}  \& {Simionescu, A.}}{{Mernier} et~al.}{2015a}]{Mernier_2015}
{Mernier} {de Plaa, J.} {Lovisari, L.} {Pinto, C.} {Zhang, Y.-Y.} {Kaastra, J. S.} {Werner, N.}  {Simionescu, A.} 2015a, \mn@doi [A\&A] {10.1051/0004-6361/201425282}, 575, A37

\bibitem[\protect\citeauthoryear{{Mernier}, {de Plaa}, {Pinto}, {Kaastra}, {Kosec}, {Zhang}, {Mao}  \& {Werner}}{{Mernier} et~al.}{2016}]{Mernier2016a}
{Mernier} F.,  {de Plaa} J.,  {Pinto} C.,  {Kaastra} J.~S.,  {Kosec} P.,  {Zhang} Y.~Y.,  {Mao} J.,   {Werner} N.,  2016, \mn@doi [\aap] {10.1051/0004-6361/201527824}, \href {https://ui.adsabs.harvard.edu/abs/2016A&A...592A.157M} {592, A157}

\bibitem[\protect\citeauthoryear{{Mernier} et~al.,}{{Mernier} et~al.}{2018a}]{Mernier2018m}
{Mernier} F.,  et~al., 2018a, \mn@doi [\mnras] {10.1093/mnrasl/sly080}, \href {https://ui.adsabs.harvard.edu/abs/2018MNRAS.478L.116M} {478, L116}

\bibitem[\protect\citeauthoryear{Mernier et~al.,}{Mernier et~al.}{2018b}]{Mernier2018c}
Mernier F.,  et~al., 2018b, \mn@doi [\mnras: Letters] {10.1093/mnrasl/sly134}, 480, L95

\bibitem[\protect\citeauthoryear{{Mernier} et~al.,}{{Mernier} et~al.}{2020}]{Mernier2020}
{Mernier} F.,  et~al., 2020, \mn@doi [Astronomische Nachrichten] {10.1002/asna.202023779}, \href {https://ui.adsabs.harvard.edu/abs/2020AN....341..203M} {341, 203}

\bibitem[\protect\citeauthoryear{Mernier et~al.,}{Mernier et~al.}{2022}]{Mernier2022b}
Mernier F.,  et~al., 2022, \mn@doi [\mnras] {10.1093/mnras/stac253}, 511, 3159

\bibitem[\protect\citeauthoryear{{Mitchell}, {Culhane}, {Davison}  \& {Ives}}{{Mitchell} et~al.}{1976}]{Mitchell1976}
{Mitchell} R.~J.,  {Culhane} J.~L.,  {Davison} P.~J.~N.,   {Ives} J.~C.,  1976, \mn@doi [\mnras] {10.1093/mnras/175.1.29P}, \href {https://ui.adsabs.harvard.edu/abs/1976MNRAS.175P..29M} {175, 29P}

\bibitem[\protect\citeauthoryear{Nomoto, Kobayashi  \& Tominaga}{Nomoto et~al.}{2013}]{Nomoto_2013}
Nomoto K.,  Kobayashi C.,   Tominaga N.,  2013, \mn@doi [\araa] {10.1146/annurev-astro-082812-140956}, 51, 457

\bibitem[\protect\citeauthoryear{Parriott \& Bregman}{Parriott \& Bregman}{2008}]{Parriott_2008}
Parriott J.~R.,  Bregman J.~N.,  2008, \mn@doi [\apj] {10.1086/588033}, 681, 1215

\bibitem[\protect\citeauthoryear{{Pl{\v{s}}ek}, {Werner}, {Grossov{\'a}}, {Topinka}, {Simionescu}  \& {Allen}}{{Pl{\v{s}}ek} et~al.}{2022}]{Plsek2022}
{Pl{\v{s}}ek} T.,  {Werner} N.,  {Grossov{\'a}} R.,  {Topinka} M.,  {Simionescu} A.,   {Allen} S.~W.,  2022, \mn@doi [\mnras] {10.1093/mnras/stac2770}, \href {https://ui.adsabs.harvard.edu/abs/2022MNRAS.517.3682P} {517, 3682}

\bibitem[\protect\citeauthoryear{{Pl{\v{s}}ek}, {Werner}, {Topinka}  \& {Simionescu}}{{Pl{\v{s}}ek} et~al.}{2023}]{Plsek2023}
{Pl{\v{s}}ek} T.,  {Werner} N.,  {Topinka} M.,   {Simionescu} A.,  2023, \mn@doi [arXiv e-prints] {10.48550/arXiv.2304.05457}, \href {https://ui.adsabs.harvard.edu/abs/2023arXiv230405457P} {p. arXiv:2304.05457}

\bibitem[\protect\citeauthoryear{Renzini \& Andreon}{Renzini \& Andreon}{2014}]{Renzini_2014}
Renzini A.,  Andreon S.,  2014, \mn@doi [\mnras] {10.1093/mnras/stu1689}, 444, 3581

\bibitem[\protect\citeauthoryear{Roediger et~al.,}{Roediger et~al.}{2015a}]{Roediger2015a}
Roediger E.,  et~al., 2015a, \mn@doi [\apj] {10.1088/0004-637x/806/1/103}, 806, 103

\bibitem[\protect\citeauthoryear{Roediger et~al.,}{Roediger et~al.}{2015b}]{Roediger2015b}
Roediger E.,  et~al., 2015b, \mn@doi [\apj] {10.1088/0004-637x/806/1/104}, 806, 104

\bibitem[\protect\citeauthoryear{{Salpeter}}{{Salpeter}}{1955a}]{Salpeter_1955}
{Salpeter} E.~E.,  1955a, \mn@doi [\apj] {10.1086/145971}, \href {https://ui.adsabs.harvard.edu/abs/1955ApJ...121..161S} {121, 161}

\bibitem[\protect\citeauthoryear{{Salpeter}}{{Salpeter}}{1955b}]{Salpater1955}
{Salpeter} E.~E.,  1955b, \mn@doi [\apj] {10.1086/145971}, \href {https://ui.adsabs.harvard.edu/abs/1955ApJ...121..161S} {121, 161}

\bibitem[\protect\citeauthoryear{{Sanders} et~al.,}{{Sanders} et~al.}{2016}]{Sanders2016}
{Sanders} J.~S.,  et~al., 2016, \mn@doi [\mnras] {10.1093/mnras/stv2972}, \href {https://ui.adsabs.harvard.edu/abs/2016MNRAS.457...82S} {457, 82}

\bibitem[\protect\citeauthoryear{Seitenzahl et~al.,}{Seitenzahl et~al.}{2013}]{Seitenzahl_2012}
Seitenzahl I.~R.,  et~al., 2013, \mn@doi [\mnras] {10.1093/mnras/sts402}, 429, 1156

\bibitem[\protect\citeauthoryear{{Serlemitsos}, {Smith}, {Boldt}, {Holt}  \& {Swank}}{{Serlemitsos} et~al.}{1977}]{Serlemitsos1977}
{Serlemitsos} P.~J.,  {Smith} B.~W.,  {Boldt} E.~A.,  {Holt} S.~S.,   {Swank} J.~H.,  1977, \mn@doi [\apjl] {10.1086/182342}, \href {https://ui.adsabs.harvard.edu/abs/1977ApJ...211L..63S} {211, L63}

\bibitem[\protect\citeauthoryear{{Simionescu} et~al.,}{{Simionescu} et~al.}{2019}]{Simionescu2019b}
{Simionescu} A.,  et~al., 2019, \mn@doi [\mnras] {10.1093/mnras/sty3220}, \href {https://ui.adsabs.harvard.edu/abs/2019MNRAS.483.1701S} {483, 1701}

\bibitem[\protect\citeauthoryear{{Smith}, {Brickhouse}, {Liedahl}  \& {Raymond}}{{Smith} et~al.}{2001}]{Smith2001}
{Smith} R.~K.,  {Brickhouse} N.~S.,  {Liedahl} D.~A.,   {Raymond} J.~C.,  2001, \mn@doi [\apjl] {10.1086/322992}, \href {https://ui.adsabs.harvard.edu/abs/2001ApJ...556L..91S} {556, L91}

\bibitem[\protect\citeauthoryear{{Sofue}}{{Sofue}}{1994}]{Sofue_1994}
{Sofue} Y.,  1994, \mn@doi [\apjl] {10.1086/187480}, \href {https://ui.adsabs.harvard.edu/abs/1994ApJ...431L..91S} {431, L91}

\bibitem[\protect\citeauthoryear{{Su}, {Irwin}, {White}  \& {Cooper}}{{Su} et~al.}{2015}]{Su_2015}
{Su} Y.,  {Irwin} J.~A.,  {White} Raymond~E. I.,   {Cooper} M.~C.,  2015, \mn@doi [\apj] {10.1088/0004-637X/806/2/156}, \href {https://ui.adsabs.harvard.edu/abs/2015ApJ...806..156S} {806, 156}

\bibitem[\protect\citeauthoryear{Su et~al.,}{Su et~al.}{2017}]{Su_2017}
Su Y.,  et~al., 2017, \mn@doi [\apj] {10.3847/1538-4357/834/1/74}, 834, 74

\bibitem[\protect\citeauthoryear{Temi et~al.,}{Temi et~al.}{2022}]{Temi_2022}
Temi P.,  et~al., 2022, \mn@doi [\apj] {10.3847/1538-4357/ac5036}, 928, 150

\bibitem[\protect\citeauthoryear{Thomas, Maraston, Schawinski, Sarzi  \& Silk}{Thomas et~al.}{2010}]{Thomas2010}
Thomas D.,  Maraston C.,  Schawinski K.,  Sarzi M.,   Silk J.,  2010, \mn@doi [\mnras] {10.1111/j.1365-2966.2010.16427.x}, 404, 1775

\bibitem[\protect\citeauthoryear{{Truong} et~al.,}{{Truong} et~al.}{2019}]{Truong2019}
{Truong} N.,  et~al., 2019, \mn@doi [\mnras] {10.1093/mnras/stz161}, \href {https://ui.adsabs.harvard.edu/abs/2019MNRAS.484.2896T} {484, 2896}

\bibitem[\protect\citeauthoryear{{Tully} et~al.,}{{Tully} et~al.}{2013}]{Tully2013}
{Tully} R.~B.,  et~al., 2013, \mn@doi [\aj] {10.1088/0004-6256/146/4/86}, \href {https://ui.adsabs.harvard.edu/abs/2013AJ....146...86T} {146, 86}

\bibitem[\protect\citeauthoryear{Urban, Werner, Allen, Simionescu  \& Mantz}{Urban et~al.}{2017}]{Urban2017}
Urban O.,  Werner N.,  Allen S.~W.,  Simionescu A.,   Mantz A.,  2017, \mn@doi [\mnras] {10.1093/mnras/stx1542}, 470, 4583

\bibitem[\protect\citeauthoryear{{Werner} \& {Mernier}}{{Werner} \& {Mernier}}{2020}]{Werner_2020}
{Werner} N.,  {Mernier} F.,  2020, in , Reviews in Frontiers of Modern Astrophysics; From Space Debris to Cosmology.
pp 279--310, \mn@doi{10.1007/978-3-030-38509-5_10}

\bibitem[\protect\citeauthoryear{{Werner}, {de Plaa, J.}, {Kaastra, J. S.}, {Vink, Jacco}, {Bleeker, J. A. M.}, {Tamura, T.}, {Peterson, J. R.}  \& {Verbunt, F.}}{{Werner} et~al.}{2006}]{Werner_2006}
{Werner} {de Plaa, J.} {Kaastra, J. S.} {Vink, Jacco} {Bleeker, J. A. M.} {Tamura, T.} {Peterson, J. R.}  {Verbunt, F.} 2006, \mn@doi [A\&A] {10.1051/0004-6361:20053868}, 449, 475

\bibitem[\protect\citeauthoryear{Werner, Urban, Simionescu  \& Allen}{Werner et~al.}{2013}]{Werner2013}
Werner N.,  Urban O.,  Simionescu A.,   Allen S.~W.,  2013, \mn@doi [Nature] {10.1038/nature12646}, 502, 656

\bibitem[\protect\citeauthoryear{Werner, McNamara, Churazov  \& Scannapieco}{Werner et~al.}{2018}]{Werner2018}
Werner N.,  McNamara B.~R.,  Churazov E.,   Scannapieco E.,  2018, \mn@doi [\ssr] {10.1007/s11214-018-0571-9}, 215, 5

\bibitem[\protect\citeauthoryear{{Willingale}, {Hands}, {Warwick}, {Snowden}  \& {Burrows}}{{Willingale} et~al.}{2003}]{Willingale2003}
{Willingale} R.,  {Hands} A.~D.~P.,  {Warwick} R.~S.,  {Snowden} S.~L.,   {Burrows} D.~N.,  2003, \mn@doi [\mnras] {10.1046/j.1365-8711.2003.06741.x}, \href {https://ui.adsabs.harvard.edu/abs/2003MNRAS.343..995W} {343, 995}

\bibitem[\protect\citeauthoryear{Zhang, Churazov, Forman  \& Jones}{Zhang et~al.}{2018}]{Zhang_2018}
Zhang C.,  Churazov E.,  Forman W.~R.,   Jones C.,  2018, \mn@doi [\mnras] {10.1093/mnras/sty2501}, 482, 20–29

\bibitem[\protect\citeauthoryear{{Zhang}, {Simionescu}, {Akamatsu}, {Kaastra}, {de Plaa}  \& {van Weeren}}{{Zhang} et~al.}{2020}]{Zhang2020}
{Zhang} X.,  {Simionescu} A.,  {Akamatsu} H.,  {Kaastra} J.~S.,  {de Plaa} J.,   {van Weeren} R.~J.,  2020, \mn@doi [\aap] {10.1051/0004-6361/202037965}, \href {https://ui.adsabs.harvard.edu/abs/2020A&A...642A..89Z} {642, A89}

\bibitem[\protect\citeauthoryear{{de Grandi} \& {Molendi}}{{de Grandi} \& {Molendi}}{2009}]{Grandi2009}
{de Grandi} S.,  {Molendi} S.,  2009, \mn@doi [\aap] {10.1051/0004-6361/200912745}, \href {https://ui.adsabs.harvard.edu/abs/2009A&A...508..565D} {508, 565}

\bibitem[\protect\citeauthoryear{{de Plaa}, {Werner}, {Bleeker}, {Vink}, {Kaastra}  \& {M{\'e}ndez}}{{de Plaa} et~al.}{2007}]{dePlaa2007}
{de Plaa} J.,  {Werner} N.,  {Bleeker} J.~A.~M.,  {Vink} J.,  {Kaastra} J.~S.,   {M{\'e}ndez} M.,  2007, \mn@doi [\aap] {10.1051/0004-6361:20066382}, \href {https://ui.adsabs.harvard.edu/abs/2007A&A...465..345D} {465, 345}

\bibitem[\protect\citeauthoryear{{de Plaa} et~al.,}{{de Plaa} et~al.}{2017a}]{dePlaa2017}
{de Plaa} J.,  et~al., 2017a, \mn@doi [\aap] {10.1051/0004-6361/201629926}, \href {https://ui.adsabs.harvard.edu/abs/2017A&A...607A..98D} {607, A98}

\bibitem[\protect\citeauthoryear{{de Plaa} et~al.,}{{de Plaa} et~al.}{2017b}]{dePlaa__2017}
{de Plaa} J.,  et~al., 2017b, \mn@doi [\aap] {10.1051/0004-6361/201629926}, \href {https://ui.adsabs.harvard.edu/abs/2017A&A...607A..98D} {607, A98}

\makeatother
\end{thebibliography}


\appendix

\section{Abundance measurement comparison of different temperature models}\label{appendix:a}

\begin{figure*}
\centering
    \includegraphics[width=0.85\textwidth]{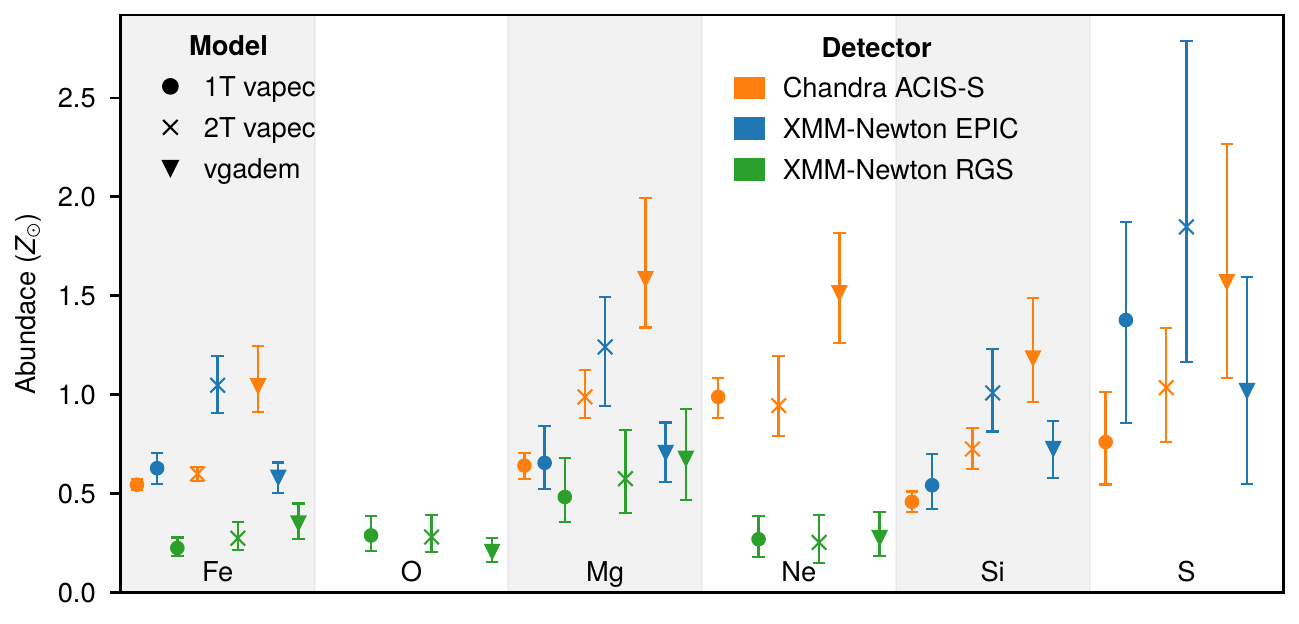}
    \includegraphics[width=0.85\textwidth]{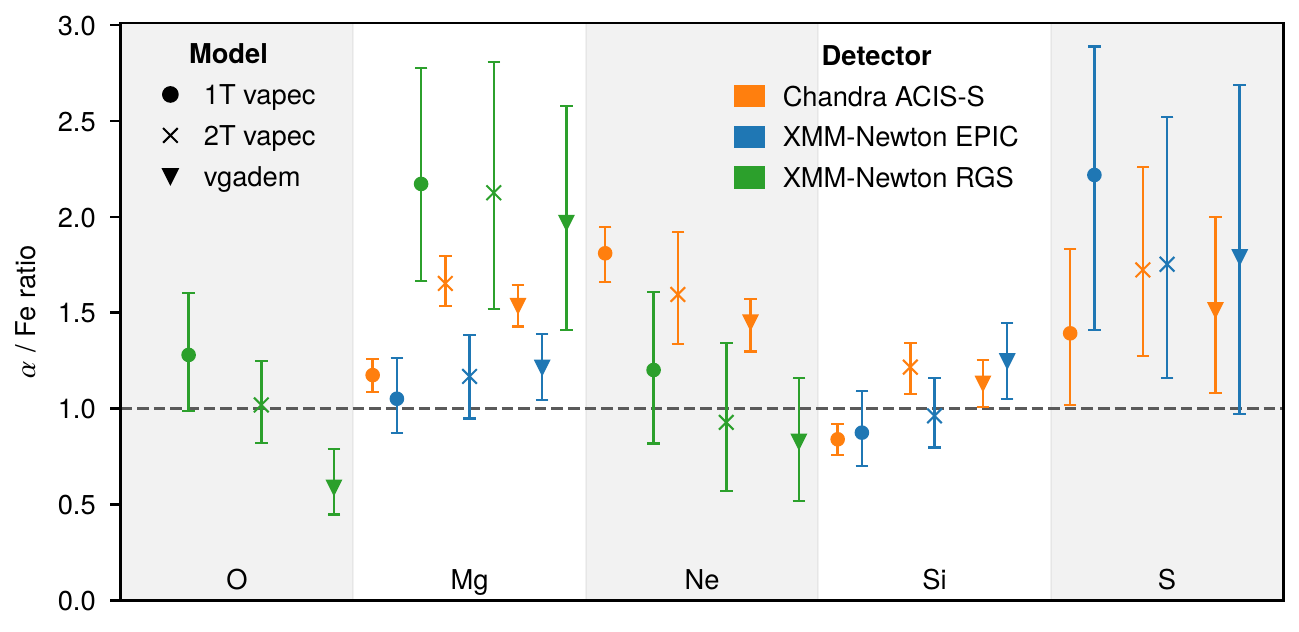}
    \caption{Abundance and ratio measurements with different temperature models using ACIS-S, EPIC and RGS data.}
    \label{fig:models_detectors}
\end{figure*}
\begin{figure*}
\centering
    \includegraphics[width=0.85\textwidth]{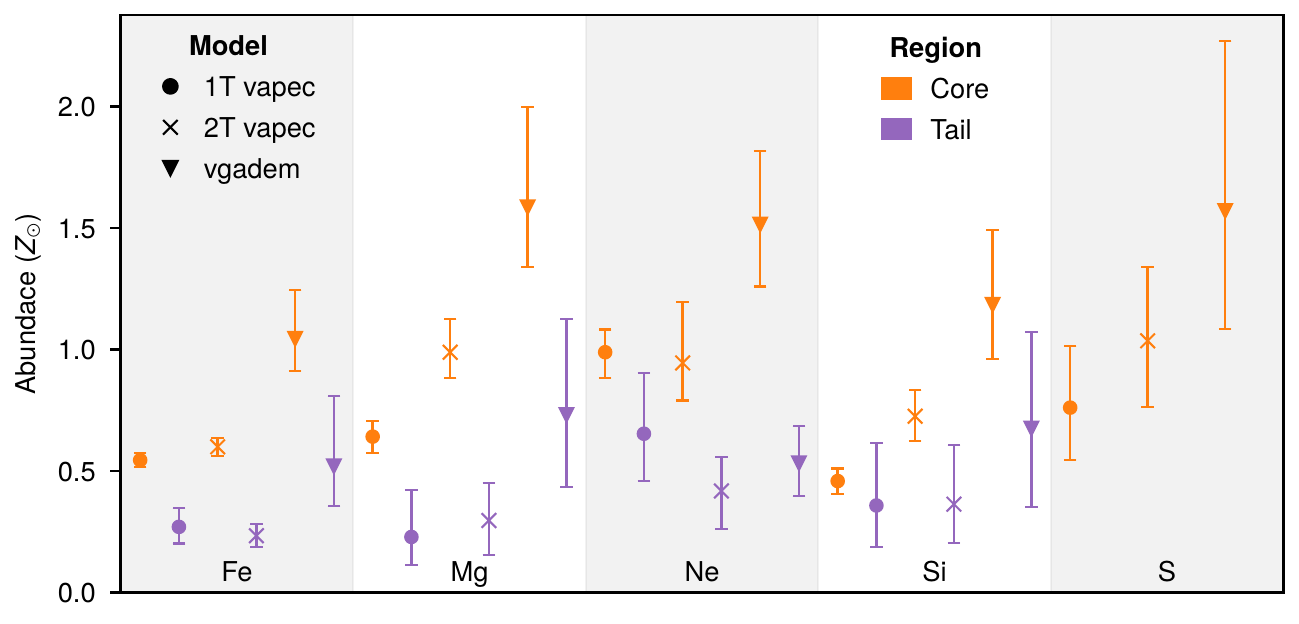}
    \includegraphics[width=0.85\textwidth]{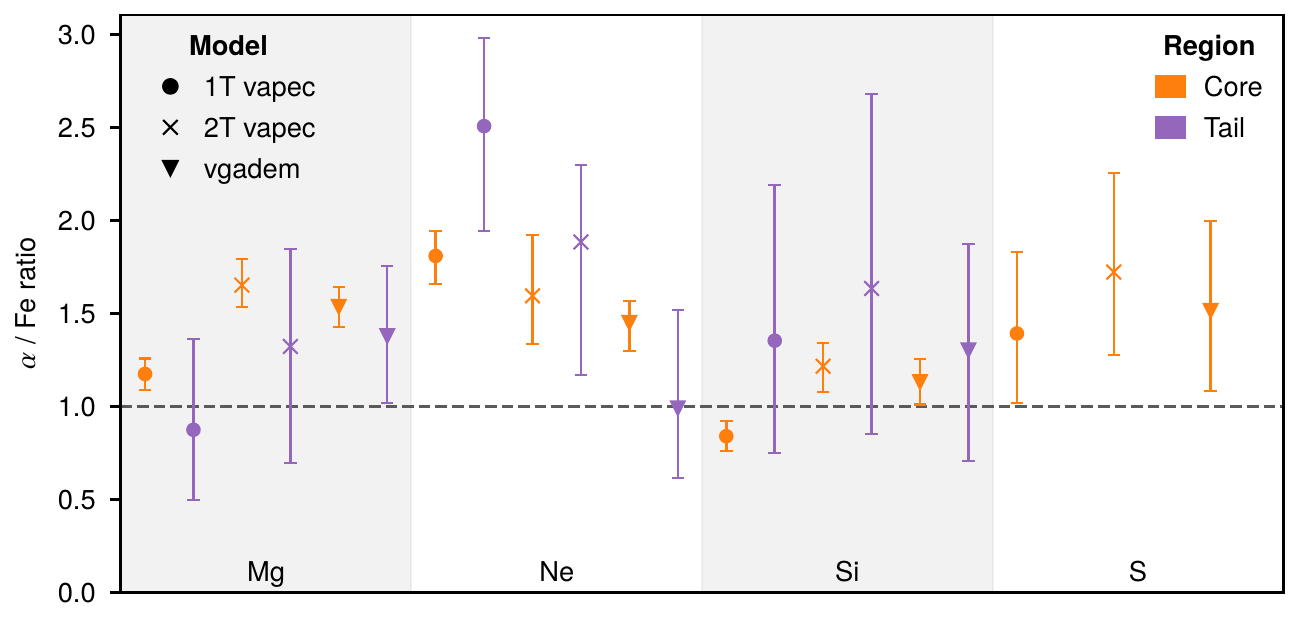}
    \caption{Abundance and ratio measurement comparison of Tail and Core regions using ACIS-S data.}
    \label{fig:40arcsec_tail}
\end{figure*}

We present the abundance and $\alpha$/Fe ratios from Table \ref{table:parameters} for comparison. In Figure \ref{fig:models_detectors}, the abundance and ratios measured with 1T \code{vapec}, 1T 2$\times$\code{vapec} and \code{vgadem} temperature models from ACIS-S, EPIC and RGS data are presented. 
Moreover, similar comparison plots of the Core and Tail regions derived with different temperature models are presented in Figure \ref{fig:40arcsec_tail}. In this plot, the derived values are from ACIS-S data.

\bsp	
\label{lastpage}
\end{document}